\documentclass[aps,prb,twocolumn,floatfix,showpacs]{revtex4-1}
\usepackage{amsmath,amssymb,graphicx,bm,color}
\def\veck{\mathbf k}
\def\vecq{\mathbf q}

\def\vecQ{\mathbf Q}

\def\be{\begin{equation}}
\def\ee{\end{equation}}

\begin{document}
\title{Strongly correlated electrons: Analytic mean-field theories with two-particle self-consistency}

\author{V\'aclav  Jani\v{s} }\author{Peter Zalom} \author{Vladislav Pokorn\'y} \author{Anton\'\i n Kl\'\i\v c} 

\affiliation{Institute of Physics, The Czech Academy of Sciences, Na Slovance 2, CZ-18221 Praha  8,  Czech Republic}

\email{janis@fzu.cz}

\date{\today}

%\maketitle

\begin{abstract}
A two-particle self-consistency is rarely part of mean-field theories. It is, however, essential for avoiding spurious critical transitions and unphysical behavior. We present a general scheme for constructing analytically controllable approximations with self-consistent equations for the two-particle vertices based on the parquet equations. We explain in detail how to reduce the full set of parquet equations not to miss quantum criticality in strong coupling. We further introduce a decoupling of  convolutions of the dynamical variables in the Bethe-Salpeter equations to make them analytically solvable. We connect the self-energy with the two-particle vertices to satisfy the Ward identity and the Schwinger-Dyson equation and discuss the role of the one-particle self-consistency in making the approximations reliable in the whole spectrum of the input parameters. Finally, we  demonstrate the general construction on the simplest static approximation that we apply to the Kondo behavior of the single-impurity Anderson model. \end{abstract}

\maketitle %newpage

\section{Introduction}
\label{sec:Intro}

Many-body systems are those in which particle interactions cannot be neglected. In particular, when they are strong, they cause critical fluctuations and may lead to qualitative changes and phase transitions in extended systems. The critical behavior must be treated self-consistently.  Even simple models of correlated elementary objects are, however, unsolvable. We hence must resort to approximations, apart from a few limiting cases where exact partial solutions exist. The approximations are either numerical or semi-analytic. The numerical approach tends to offer unbiased approximations with all degrees of freedom left in play, while the analytic approximations are based on a reduction of complexity of interaction effects.  The numerical calculations offer  good quantitative predictions  that can set trends in the dependence of the solution on the  model input parameters. The latter schemes aspire to reproduce qualitative features of the exact solution. They, unlike the numerical methods,  can address and control  singularities directly.       

The first successful effort to tame the non-analytic behavior at the critical point of the continuous phase transitions in classical statistical models is the Landau mean-field theory in which the free energy is expanded in the small order parameter around the critical point  \cite{Landau:1937}. This simplest local and static self-consistent approximation inspired further attempts to improve upon it by including dynamical corrections \cite{Brout:1959aa,Brout:1960aa,Horwitz:1961aa}. But these  attempts failed to match consistently ordered and disordered phases  at the critical point \cite{Englert:1963aa}. These inconsistencies were removed later with a quite different improvement of the mean field by using scaling arguments and the renormalization group \cite{Wilson:1971aa,Wilson:1971ab}. A mean-field approximation remains nevertheless the starting point also for the renormalization-group construction in analytic treatments.

The concept of a local, comprehensive mean-field theory was revitalized by realizing that it can be obtained as an exact solution in the limit of high spatial dimensions for classical spin models \cite{Brout:1960aa,Fisher:1964aa,Thompson:1974aa,Sherrington:1975aa},  as well as for quantum itinerant models \cite{Metzner:1989aa,Janis:1989aa,Brandt:1989aa,Janis:1991aa}.  In particular, the limit to high spatial dimensions in models of correlated and disordered electrons  initiated a boom in the applications of the dynamical mean-field theory (DMFT) \cite{Georges:1996aa,Kotliar:2006aa}.  The major asset of the DMFT is its unbiased inclusion of quantum fluctuations missed in classical static, weak-coupling  theories. It hence offers a reliable way to investigate the strong-coupling limit of correlated electron systems at low temperatures. Since the single-impurity Anderson model (SIAM) is contained as the first, non-self-consistent iteration to the DMFT, the advanced methods of SIAM have been used to derive impurity solvers for the  the dynamical mean-field approximations.

The standard mean-field theories, including DMFT, introduce renormalizations only for one-particle quantities represented by order parameters or the self-energy.  They contain no renormalizations of vertex functions and hence, there is no direct control of the singularities in the Bethe-Salpeter equations for the two-particle response functions. The attempts to include two-particle and vertex renormalizations in the perturbation expansion are presently mostly made via non-local corrections to the DMFT \cite{Rohringer:2018aa} or loop corrections with the functional renormalization group approach \cite{Kopietz:2010ab}. Recently, a two-particle self-consistency resulting from the parquet approach was used to construct an analytic mean-field theory for strong coupling \cite{Janis:2007aa,Janis:2008ab,Janis:2017aa,Janis:2017ab}. The solution was shown to reproduce qualitatively correctly the Kondo effect in the strong-coupling limit of SIAM at zero temperature.  It can be viewed upon as a consistent generalization of the Hartree approximation to strong coupling which is free of the spurious transition to an ordered phase of the weak-coupling solution. An effective interaction, as the only two-particle object determined self-consistently, is a static approximation to the irreducible vertex in the electron-hole scattering channel which resembles the GW construction \cite{Hedin:1965aa,Aryasetiawan:1997aa}.  

Although popularity of two-particle approaches has increased in recent years \cite{Janis:1998aa,Janis:1999aa,Toschi:2007aa,Rubtsov:2008aa,Rohringer:2011aa,Rubtsov:2012aa,Rohringer:2013aa,Valli:2015aa,Hirschmeier:2015aa,Ayral:2016aa,Kugler:2018aa,Del_Re:2019aa}, complexity of two-particle vertices demands the application of heavy numerics to reach quantitative results unless further approximations are used \cite{Yang:2009aa,Tam:2013aa,Li:2016aa}.  The numerical solutions do not allow for the identification  of the relevant degrees of freedom and for the control of the non-analytic behavior in the critical regions of vertex functions offered by mean-field theories. One has to introduce specific simplifications making the use of two-particle functions in the construction of mean-field approximations effective. 
  
A few attempts were also made to simplify approximations for two-particle functions and to derive a  set of static effective interactions determined self-consistently \cite{Vilk:1997aa} and \cite{Kusunose:2010aa}. The former approach determines effective interactions in the charge and magnetic channels from local sum rules, while the latter determines the irreducible vertices in all two-particle channels from the crossing symmetry. Neither of these approaches was able to determine the Kondo scale in SIAM analytically.
 
The approximate scheme developed in Refs.~\cite{Janis:2007aa,Janis:2008ab,Janis:2017aa,Janis:2017ab} used  three leading principles of simplifying the equations for the two-particle vertices to end up with an analytic, mean-field-like theory of quantum criticality.  1) The full scheme of the parquet equations was replaced by a reduced set, 2) the dependence of the irreducible vertex on its dynamical variables was suppressed, and 3) the standard construction of approximate theories of Baym and Kadanoff \cite{Baym:1961aa,Baym:1962aa} had to be left to comply with the Ward identity and to keep the approximation conserving whereby two self-energies were introduced. These steps were performed pragmatically, goal-directed without fully clarifying their general meaning for the construction of analytic approximations with the self-consistent determination of two-particle vertex functions.    

The aim of this paper is to present a systematic derivation of mean-field theories with a two-particle self-consistency  based on the parquet equations with the necessary simplifications leading to semi-analytic conserving approximations. The resulting approximations are free of unphysical behavior and spurious phase transitions and are applicable in strong coupling, both in disordered as well as in ordered phases of models of correlated electrons. In particular, we explain why a reduction of the full set of parquet equations with the bare interaction as the fully two-particle irreducible vertex is needed to reach the quantum critical behavior in the Bethe-Salpeter equations. We further explain how to separate the relevant from irrelevant dynamical fluctuations near the singularities in the two-particle vertex and how to decouple convolutions of fermionic Matsubara frequencies in the Bethe-Salpeter equations to make them analytically solvable. But most importantly, we give the proper meaning to two self-energies understood now as parts of  a single self-energy with even and odd symmetry with respect to the symmetry-breaking field controlling the critical behavior. The former self-energy is determined from the dynamical Schwinger-Dyson equation. The latter is coupled with the electron-hole irreducible vertex via the Ward identity linearized in the symmetry-breaking field. It plays the role of the order parameter. The full self-energy is then compatible with both, the Ward identity and the Schwinger-Dyson equation. We thus set a framework for systematic improvements of the impurity solver from Refs.~\cite{Janis:2007aa,Janis:2008ab,Janis:2017aa,Janis:2017ab} to dynamical mean-field-like approximations with a two-particle self-consistency offering a reliable description of thermodynamic and spectral properties in the strong-coupling limit in different settings of impurity and bulk models of correlated electrons.   

\section{Two-particle self-consistency: Reduced parquet equations}
\label{sec:2PSC-PE}

A controlled and reliable way to suppress spurious transitions of the weak-coupling mean-field approximation is to introduce a two-particle self-consistency where two-particle vertices are determined self-consistently from non-linear equations. Only integrable singularities survive there as real phase transitions.  One possibility to reach a two-particle self-consistency is to replace the bare interaction in response functions by effective ones determined self-consistently \cite{Vilk:1997aa}. A more systematic way to do this is to use the parquet equations introduced to condensed matter by De Dominicis and Martin \cite{DeDominicis:1964aa,DeDominicis:1964ab}. The full set of parquet equations was used to solve the soft X-ray problem \cite{Roulet:1969aa},  to understand the local-moment formation \cite{Weiner:1970aa,Weiner:1971aa} and was also applied to SIAM \cite{Bickers:1991aa,Bickers:1992aa}. Unfortunately, the parquet equations simultaneously self-consistent at the one and two-particle level have not brought much progress beyond the fluctuation exchange, where only one-particle self-consistency is kept \cite{Bickers:1989ab}. Although the spurious phase transition of the  Hartree theory was suppressed with the one-particle self-consistency no Kondo limit was reproduced with it \cite{Hamann:1969aa,Chen:1992aa}. There is a general problem with the full set of parquet equations with the bare interaction as the completely irreducible vertex. Quantum criticality and the Kondo effect in SIAM are completely missed \cite{Janis:2006ab}. One must either replace the bare interaction by a more complex vertex or reduce appropriately the parquet equations to be able to reach the Kondo limit.  Our aim is to find an analytically tractable  mean-field theory for strongly correlated electron systems for which purpose we reduce the parquet equations and simplify their complexity.

\subsection{Reduction scheme of the parquet equations}
\label{sec:ReductionScheme}

\subsubsection{Two-channel parquet equations}
\label{sec:2CPE}

The full parquet scheme contains three scattering channels represented by three different Bethe-Salpeter equations for the full two-particle vertex. They contain sums of multiple scatterings of singlet electron-hole ($eh$) pairs, electron-electron pairs ($ee$), and triplet electron-hole pairs ($\overline{eh}$) \cite{Bickers:1991ab}.  It is, however, sufficient to use only the singlet electron-electron and electron-hole multiple scatterings, since the third, $\overline{eh}$ channel, shares the critical fluctuations with either of the two fundamental channels \cite{Janis:1998aa}. We use the Hubbard model and SIAM to study the strong-coupling limit in correlated electron systems. The critical behavior in strong coupling is not by this specification qualitatively affected.  

The Bethe-Salpeter equation for the full singlet two-particle vertex $\Gamma_{\sigma\bar{\sigma}}$ with the irreducible vertex $ \Lambda^{eh}_{\sigma\bar{\sigma}}$ and $\bar{\sigma} = - \sigma$ in the electron-hole channel is
\begin{widetext}
\begin{multline}\label{eq:Gamma-eh}
 \Gamma_{\sigma\bar{\sigma}}(\veck,i\omega_{n},\veck',i\omega_{n'}; \vecq, i\nu_{m}) 
 = \Lambda^{eh}_{\sigma\bar{\sigma}}(\veck,i\omega_{n},\veck',i\omega_{n'}; \vecq, i\nu_{m})  - \frac 1N\sum_{\veck''}\frac 1\beta\sum_{\omega_{l}} \Lambda^{eh}_{\sigma\bar{\sigma}}(\veck,i\omega_{n},\veck'',i\omega_{l};\vecq,  i\nu_{m})  
 \\
 \times G_{\sigma}(\veck'',i\omega_{l}) G_{\bar{\sigma}}(\veck'' + \vecq, i \omega_{m + l}) \Gamma_{\sigma\bar{\sigma}}(\veck'',i\omega_{l},\veck', i\omega_{n'};\vecq,  i\nu_{m}) \,,
 \end{multline}
 \end{widetext}
where $N$ is the number of lattice sites, $\beta = 1/k_{B}T$ and $G_{\sigma}$ is the  propagator of the electron with spin $\sigma$.

Analogously the Bethe-Salpeter equation for the same vertex in the electron-electron channel is 
\begin{widetext}
\begin{multline}\label{eq:Gamma-ee}
 \Gamma_{\sigma\bar{\sigma}}(\veck,i\omega_{n},\veck',i\omega_{n'}; \vecq, i\nu_{m}) 
 = \Lambda^{ee}_{\sigma\bar{\sigma}}(\veck,i\omega_{n},\veck',i\omega_{n'}; \vecq, i\nu_{m})  - \frac 1N\sum_{\veck''}\frac 1\beta\sum_{\omega_{l}} \Lambda^{ee}_{\sigma\bar{\sigma}}(\veck,i\omega_{n},\veck'',i\omega_{l};\vecq + \veck'- \veck'',  i\nu_{m + n'- l})  
 \\
 \times G_{\sigma}(\veck'',i\omega_{l}) G_{\bar{\sigma}}(\vecq + \veck + \veck'- \veck'', i \omega_{m +n + n'- l}) \Gamma_{\sigma\bar{\sigma}}(\veck'',i\omega_{l},\veck', i\omega_{n'};\vecq + \veck - \veck'',  i\nu_{m + n - l})\,.
 \end{multline} 
%\end{widetext}
%
If we introduce a fully two-particle irreducible vertex $I_{\sigma\bar{\sigma}}$ we can use the parquet decomposition of the full vertex 
%\begin{widetext}
\begin{multline}\label{eq:Parquet-fundamental}
\Gamma_{\sigma\bar{\sigma}}(\veck,i\omega_{n},\veck',i\omega_{n'}; \vecq, i\nu_{m}) = \Lambda^{eh}_{\sigma\bar{\sigma}}(\veck,i\omega_{n},\veck',i\omega_{n'}; \vecq, i\nu_{m}) + \Lambda^{ee}_{\sigma\bar{\sigma}}(\veck,i\omega_{n},\veck',i\omega_{n'}; \vecq, i\nu_{m}) 
\\
- I_{\sigma\bar{\sigma}}(\veck,i\omega_{n},\veck',i\omega_{n'}; \vecq, i\nu_{m}) \,.
\end{multline}
\end{widetext}
The  parquet decomposition holds if the set of \textit{reducible} diagrams in the electron-hole channel has no overlap with the set of \textit{reducible} diagrams in the electron-electron channel \cite{Janis:2009aa}. The two-particle self-consistency is then obtained by replacing the full vertex $\Gamma_{\sigma\bar{\sigma}}$  by the above parquet decomposition  in the Bethe-Salpeter equations~\eqref{eq:Gamma-eh} and~\eqref{eq:Gamma-ee} to obtain a set of self-consistent equations for the irreducible vertices $\Lambda^{eh}$ and  $\Lambda^{ee}$.  One standardly chooses the bare Hubbard interaction $U$ as the vertex irreducible in both two-particle scattering channels.
\begin{table}[h!]
    \caption{Notation for 2P vertices used here and in our preceding papers  compared to that used in recent papers on parquet equations  as in review~\cite{Rohringer:2018aa}, Ref.~\cite{Bickers:1991ab}, and one of the first papers on parquet equations in condensed mater~\cite{Roulet:1969aa}.  We skipped the spin indices but kept the channel index $\alpha = eh, ee, \overline{eh}$. \label{tab:table1}}
  \begin{center}
    \begin{tabular}{l|c|c|c|c} % <-- Changed to S here.
    \hline\hline
    \textbf{Source}  &\textbf{Completely} & \textbf{Channel} & \textbf{Channel} &\textbf{Full}\\
       &  irreducible &  irreducible & reducible & vertex \\
      \hline
      This paper & $I$ & $\Lambda^{\alpha}$& $K^{\alpha}$  & $\Gamma$\\
      Ref.~\cite{Rohringer:2018aa} & $\Lambda$ & $\Gamma^{\alpha}$ & $\Phi^{\alpha}$ & $F$\\
      Ref.~\cite{Bickers:1991ab} & $\Lambda^{irr}$ & $ \Gamma_{\alpha}$ & --  & $\Gamma$\\
      Ref.~\cite{Roulet:1969aa} & $R$ & $ I^{\alpha}$ & $\gamma^{\alpha}$ & $\Gamma$\\
 \hline\hline 
    \end{tabular}

  \end{center}
\end{table}
We kept notation $\Gamma$ for the full vertex and used symbol $\Lambda^{\alpha}$ for the irreducible vertex in the channel $\alpha$. Recent publications on the parquet equations as reviewed in Ref.~\cite{Rohringer:2018aa} used a slightly different notation. We related these two notations in Table~\ref{tab:table1} and compared them also with that in a seminal paper on the parquet equations, Ref.~\cite{Bickers:1991ab} and with that in an early application of the parquet equations in condensed matter \cite{Roulet:1969aa}.  

 %\vspace{12pt}

\subsubsection{Critical region of the two-particle vertex}
\label{sec:2Pcriticality}

One of the Bethe-Salpeter equations approaches a singularity, a divergence in the full vertex at the critical point, when we increase the particle interaction. It is Eq.~\eqref{eq:Gamma-eh} for the magnetic systems (repulsive interaction) and Eq.~\eqref{eq:Gamma-ee} for the superconducting systems (attractive interaction). This can be seen from the weak-coupling approximation in the Bethe-Salpeter equation where the irreducible vertex $\Lambda$ is replaced by the bare interaction. We assume that this critical behavior can analytically be continued to strong coupling where the bare interaction must be renormalized and replaced by vertex $\Lambda$. It means that the divergence in the two-particle vertex emerges in the limit $\vecq\to q_{0}$ and $\nu_{m}\to 0$ and the irreducible vertex in the singular Bethe-Salpeter equation remains bounded and non-singular in the low-energy limit $\vecq\to\vecq_{0}$ and $\nu_{m}\to 0$ \footnote{The high-frequency structure  of the irreducible vertex may be rather complex with non-analyticities \cite{Janis:2014aa,Schafer:2013aa,Chalupa:2018aa}. They are related to increasing imaginary part of the self-energy and are precursors of the metal-insulator transition. They may affect the low-frequency behavior only beyond the Fermi-liquid regime, since the imaginary part of the self-energy vanishes at the Fermi energy for Fermi liquids.}. We will explicitly consider the magnetic case with the repulsive interaction and criticality in Bethe-Salpeter equation~\eqref{eq:Gamma-eh}.     

We can single out the relevant fluctuations in the critical region in Eq.~\eqref{eq:Gamma-ee} in the spirit of the renormalization group. The relevant fluctuations are those that make the dominant contribution to vertex $ \Lambda_{\sigma\bar{\sigma}}^{ee}(\veck,i\omega_{n}, \veck'',  i\omega_{l} ; \vecq + \veck' - \veck'',  i\nu_{m + n' - l} )$.  They are controlled by the transfer momentum $\vecq+ \veck' - \veck'' \to \vecq_{0}$ and frequency $i\nu_{m + n'- l} \to 0$.   For simplicity we assume a homogeneous order and choose $\vecq_{0}=\mathbf{0}$. The fluctuations in the fermionic variables remain irrelevant in the critical region and can be neglected. Equation~\eqref{eq:Gamma-ee} in the critical region then reduces to   
\begin{widetext}
\begin{multline}\label{eq:Lambda-eh}
 \Lambda_{\sigma\bar{\sigma}}^{eh}(\veck,i\omega_{n},\veck',i\omega_{n'}; \vecq, i\nu_{m}) 
 = U  - \frac 1N\sum_{\vecQ}\frac 1\beta\sum_{\nu_{l}} \Lambda^{ee}_{\sigma\bar{\sigma}}(\veck,i\omega_{n},\veck' + \vecq + \vecQ ,i\omega_{n' + m + l}; - \vecQ  - i\nu_{l})  
 \\
 \times G_{\sigma}(\veck'+ \vecq + \vecQ,i\omega_{n' + m + l}) G_{\bar{\sigma}}(\veck - \vecQ, i \omega_{n - l}) \Gamma_{\sigma\bar{\sigma}}(\veck'+ \vecq +\vecQ,i\omega_{n'+m + l},\veck', i\omega_{n'};\veck - \veck'-\vecQ,  i\nu_{n - n'- l})\,,
 \end{multline} 
\end{widetext}
where we used the parquet equation~\eqref{eq:Parquet-fundamental} with the bare interaction as the completely irreducible vertex, $I_{\sigma\bar{\sigma}} =U$.

\subsubsection{Reduction of parquet equations}
\label{sec:ReductionParquet}
 
The assumption about the possibility to follow continuously the critical behavior from weak to strong coupling demands that the irreducible vertex remains free of divergences in the low-energy limit. However, equation~\eqref{eq:Lambda-eh} may lead to divergences in $\Lambda_{\sigma\bar{\sigma}}^{eh}$ when $\veck - \veck' =\mathbf{0}$ and $\omega_{n} - \omega_{n'} =0$. If there is no correction to the bare interaction, the solution of this equation never reaches a critical behavior. The irreducible vertex would become also divergent which is incompatible with the two-particle self-consistency in the parquet equations. The critical behavior would be fully suppressed. To preserve the critical behavior one should extend the two-channel approximation by including  the third channel and by replacing the bare interaction by a more complex completely irreducible vertex that would lead to the cancellation of the singular contributions to       
$\Lambda_{\sigma\bar{\sigma}}^{eh}$ \cite{Janis:2006ab}. This is, however, a tremendous task that would prevent reaching the desired objective of the analytic control of the critical behavior. 

Alternatively, one can resort to a ``poor-man approach'' and keep the critical behavior of the non-renormalized theory by removing  the super-divergent term from the convolution of two divergent vertices in Eq.~\eqref{eq:Lambda-eh}. This leads then to a reduction of the parquet equations. This is achieved by the following replacement of the convolution on the right-hand side of   Eq.~\eqref{eq:Lambda-eh}   
\begin{widetext}
\begin{multline}
\Lambda^{ee}_{\sigma\bar{\sigma}}(\veck,i\omega_{n},\veck'',i\omega_{l};\vecq' - \veck - \veck'',  i\nu_{m' -n - l})  
 G_{\sigma}(\veck'',i\omega_{l}) G_{\bar{\sigma}}(\vecq' - \veck'', i \omega_{m' - l})
 \Gamma_{\sigma\bar{\sigma}}(\veck'',i\omega_{l},\veck', i\omega_{n'};\vecq' - \veck' - \veck'',  i\nu_{m' - n'- l})  
 \\
 \to K_{\sigma\bar{\sigma}}(\veck,i\omega_{n},\veck'',i\omega_{l};\vecq' - \veck - \veck'',  i\nu_{m' - n - l})  G_{\sigma}(\veck'',i\omega_{l}) G_{\bar{\sigma}} (\vecq' - \veck'', i \omega_{m' - l})\Lambda_{\sigma\bar{\sigma}}(\veck'',i\omega_{l},\veck', i\omega_{n'};\vecq' - \veck' - \veck'',  i\nu_{m' -n' - l})
\,,
\end{multline}
\end{widetext}
 where we denoted $K_{\sigma\bar{\sigma}} = \Gamma_{\sigma\bar{\sigma}} - \Lambda^{eh}_{\sigma\bar{\sigma}}$, the reducible vertex in the electron-hole channel with $\Lambda = \Lambda^{eh}$ and used $\veck''= \veck'+ \vecq + \vecQ$, $\vecq'= \vecq + \veck + \veck'$, $m'= m + n + n'$.    
Since the completely irreducible vertex is a static constant, the irreducible vertex $ \Lambda_{\sigma\bar{\sigma}}^{eh}(\veck,i\omega_{n},\veck',i\omega_{n'}; \vecq, i\nu_{m}) $ depends only on fermionic variables $\veck'+ \vecq$, $i\omega_{n' + m}$ and $\veck$, $i\omega_{n}$. We redefine the transfer momentum in the irreducible vertex $\Lambda_{\sigma\bar{\sigma}}(\veck,\omega_{n}; \veck', i\omega _{n'};\vecq,i\nu_{m}) \to \Lambda_{\sigma\bar{\sigma}}(\veck,i\omega_{n},\vecq + \veck', i\omega_{m + n'})$ to simplify the dependence of the dynamical variables. 
%This transformation is graphically represented in Fig.~\ref{fig:Lambda-trans}.
%%
%\begin{figure}.
%\includegraphics[width=8cm]{Lambda_transform}
%
%\caption{Transformation of momentum and frequency dependence of the irreducible vertex $\Lambda$.  We used a four-vector notation, $k =(\veck,i\omega_{n})$, $q =(\vecq,i\nu_{m})$, and $Q = k + k'+ q$. \label{fig:Lambda-trans}}
%\end{figure}

\begin{figure}
\includegraphics[width=8cm]{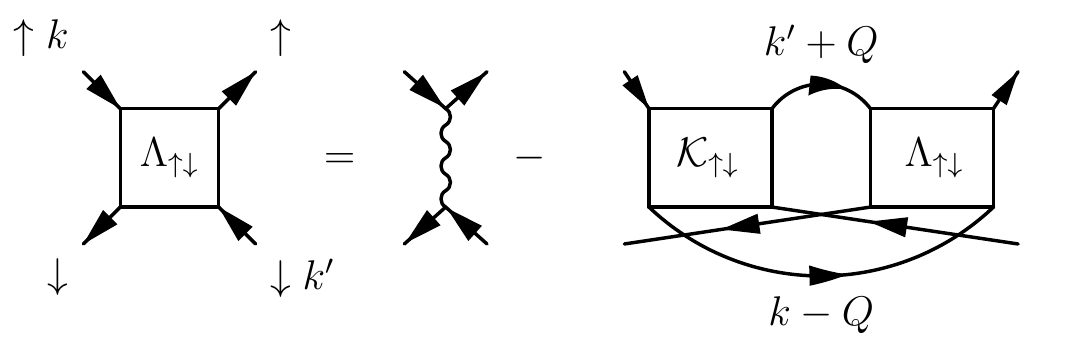}
\caption{The Bethe-Salpeter equation~\eqref{eq:Lambda-reduced-bar} for the irreducible vertex $\Lambda_{\uparrow\downarrow}$ with the integral kernel, the reducible vertex  $\mathcal{K}_{\uparrow\downarrow}$ from the the electron-hole channel. We used the four-vector notation, $k =(\veck,i\omega_{n})$ for the fermionic  and $Q= (\vecQ,i\nu_{m})$ for the bosonic variables. Both sides have the same external labels and $Q$ is the internal variable. The vertical wavy line is the bare Hubbard interaction.  \label{fig:RPE-ee} }
\end{figure}

The irreducible vertex is then determined from an integral  equation
\begin{widetext}
\begin{multline}\label{eq:Lambda-reduced-bar}
 \Lambda_{\sigma\bar{\sigma}}(\veck,i\omega_{n};\veck',  i\omega_{n'}) = U 
 - \frac 1N\sum_{\vecQ}  \frac 1\beta\sum_{\nu_{l}} K_{\sigma\bar{\sigma}}(\veck,i\omega_{n}, \veck' + \vecQ,  i\omega_{n'} + i\nu_{l}; -\vecQ, - i\nu_{l} )
 \\
 \times  G_{\sigma}( \veck' + \vecQ,  i\omega_{n'} + i\nu_{l}) G_{\bar{\sigma}} (\veck -\vecQ , i\omega_{n} - i\nu_{l})  \Lambda_{\sigma\bar{\sigma}}(\veck' + \vecQ ,i \omega_{n'} + i\nu_{l },\veck - \vecQ, i\omega_{n} -  i\nu_{l})\,.
 \end{multline} 
\end{widetext}
Its diagrammatic representation is plotted in Fig.~\ref{fig:RPE-ee}.
\begin{figure}
\includegraphics[width=8.5cm]{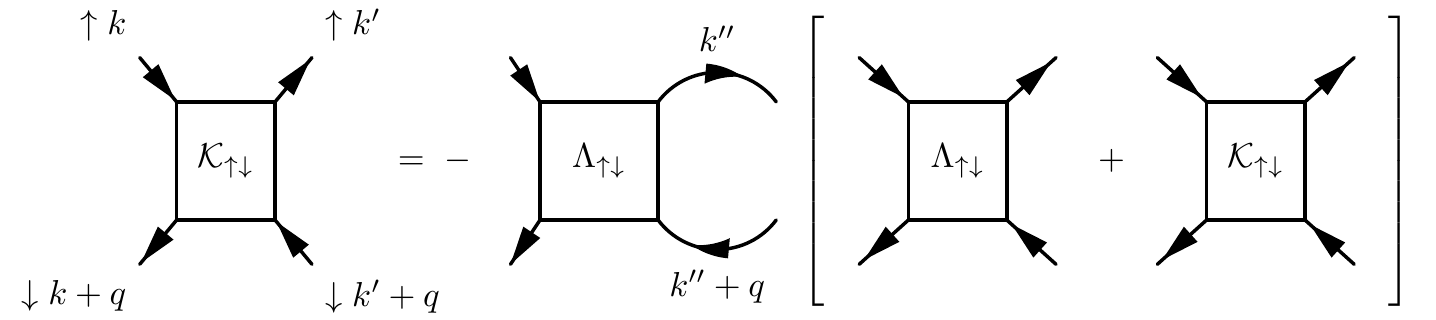}
\caption{The reduced Bethe-Salpeter equation~\eqref{eq:K-reduced-bar} for the reducible vertex $\mathcal{K}_{\uparrow\downarrow}$  from the the electron-hole channel.  Both sides have the same external labels and $k''$ is the internal (integration) variable. \label{fig:RPE-eh} }
\end{figure}
There is no change in the critical Bethe-Salpeter equation~\eqref{eq:Gamma-eh}. The reducible vertex then is 
 \begin{widetext}
 \begin{multline}\label{eq:K-reduced-bar}
 K_{\sigma\bar{\sigma}}(\veck,i\omega_{n},\veck',i\omega_{n'}; \vecq, i\nu_{m}) 
 = - \frac 1N\sum_{\veck''}\frac 1\beta\sum_{\omega_{l}} \Lambda_{\sigma\bar{\sigma}}(\veck,i\omega_{n};\vecq + \veck'',  i\omega_{m + l}) G_{\bar{\sigma}} (\veck'' + \vecq, i \omega_{m + l})
 \\
 \times   G_{\sigma}(\veck'',i\omega_{l})\left[ K_{\sigma\bar{\sigma}}(\veck'',i\omega_{l},\veck', i\omega_{n'};\vecq,  i\nu_{m}) 
 %\right. \\ \left.
 + \Lambda_{\sigma\bar{\sigma}}(\veck'',i\omega_{l};\vecq + \veck',i\omega_{m + n'}) \right] 
 \end{multline}
 \end{widetext}
 with its diagrammatic representation in Fig.~\ref{fig:RPE-eh}.

Equations~\eqref{eq:Lambda-reduced-bar} and~\eqref{eq:K-reduced-bar} form the set of the reduced parquet equations that introduce a two-particle self-consistency allowing for the extension of the critical behavior from weak-coupling  approximations continuously to strong coupling.   The reduced parquet equations~\eqref{eq:Lambda-reduced-bar}-\eqref{eq:K-reduced-bar} are the starting point for the investigation of the critical behavior of the two-particle vertex.

\subsubsection{Low momentum and frequency limit}
\label{sec:Low-energy-limit}

The reduced parquet equations~\eqref{eq:Lambda-reduced-bar}-\eqref{eq:K-reduced-bar} in their general form do not allow for analytic continuation to real frequencies with contour integrals involving Fermi and Bose distributions. To reach the goal of analytic control, one has to resort to specific limits separating the fermionic and bosonic degrees of freedom. We will consider two cases of the low-momentum and  low-frequency limit of external variables of the two-particle vertices, $\vecq,\veck,\veck'\to \mathbf{0}$ and $\nu_{m,}\omega_{n},\omega_{n'} \to 0$, where we can obtain explicit solutions of the reduced parquet equations. The two cases differ in the ratio of the bosonic and fermionic variables. If we assume $|\veck|/|\vecq|, |\veck'|/|\vecq| \to 0$ and simultaneously $\omega_{n}/\nu_{m}, \omega_{n'}/\nu_{m} \to 0$ we reproduce the static approximation introduced and analyzed in Refs.~\cite{Janis:2017aa,Janis:2017ab}. This approximation works well only at zero temperature and in the spin-symmetric state. To cover also non-zero temperatures and the ordered state we have to consider the opposite ratio,  $|\vecq|/|\veck|, |\vecq|/|\veck'| \to 0$ with $k,k'\approx k_{F}$ and $\nu_{m}/\omega_{n}, \nu_{m}/\omega_{n'} \to 0$. We will now turn the the latter case.  

Since the irreducible vertex $\Lambda_{\uparrow\downarrow}(\veck'+ \vecQ,i\omega_{n'+ l}; \veck - \vecQ,  i\omega_{n-l})$  in Eq.~\eqref{eq:Lambda-reduced-bar}is bounded in the critical region and only small values of momentum $\vecQ$ and frequency $\nu_{l}$ are relevant,  we can neglect its fluctuations. We then obtain an explicit equation for the irreducible verrtex
\begin{widetext}
  \begin{multline}\label{Lambda-Matsubara}
 \Lambda_{\uparrow\downarrow}(\veck,i\omega_{n};\veck',  i\omega_{n'})
 \\
  = \frac U{1 + N^{-1}\sum_{\vecQ}\beta^{-1}\sum_{\nu_{m}} K_{\uparrow\downarrow}(\veck',i\omega_{n'},\veck,i\omega_{n};-\vecQ, -i\nu_{m})G_{\uparrow}(\veck+ \vecQ,i\omega_{n+ m}) G_{\downarrow}(\veck' - \vecQ,i\omega_{n' - m}) } \,.
 \end{multline}
%\end{widetext}
%
The equation for the reducible vertex for small transfer momentum and frequency in the critical region reads as
%
%\begin{widetext}
\begin{multline}\label{kappa-general}
\frac 1N\sum_{\veck''} \frac 1\beta\sum_{\omega_{l}}\left[\beta N \delta_{\veck,\veck''}\delta_{n,l} + \Lambda_{\sigma\bar{\sigma}}(\veck,i\omega_{n},\veck'' + \vecq,  i\nu_{m} + i\omega_{l})  G_{\bar{\sigma}} (\vecq + \veck'',i\nu_{m} + i\omega_{l})G_{\sigma}(\veck'', i\omega_{l}) \right]
\\
\times K_{\sigma\bar{\sigma}}(\veck'',i\omega_{l},\veck',i\omega_{n'};\vecq,i\nu_{m})  = - \frac 1N\sum_{\veck''}\frac 1\beta\sum_{\omega_{l}} \Lambda_{\sigma\bar{\sigma}}(\veck,i\omega_{n};\veck'', i\omega_{l})  G_{\bar{\sigma}} (\veck'',i\omega_{l}) G_{\sigma}(\veck'',i\omega_{l})
\Lambda_{\sigma\bar{\sigma}}(\veck'',i\omega_{l},\veck',i\omega_{n'})  \,.
\end{multline}
\end{widetext}
We neglected the dependence of the sum on the right-hand side of Eq.~\eqref{kappa-general} on the transfer momentum $\vecq$ and frequency $\nu_{m}$. It is irrelevant for the critical behavior of the reducible vertex. Equation~\eqref{kappa-general} cannot, however, be solved analytically and further approximations are needed. We do it in the following section~\ref{sec:MF-general}.

%The reason for the solution of  the full parquet equations to suppress the critical point is the existence of a super-divergent term containing a convolution of two originally non-divergent vertices turning divergent by the two-particle self-consistency, $\Lambda^{ee}_{\sigma\bar{\sigma}}$ in the magnetic systems. This super-divergent term is compensated in the exact solution by contributions from dynamical corrections to the fully irreducible vertex.  A way to reach the critical region of the singularity in the Bethe-Salpeter equations with the bare interaction as the fully irreducible vertex is to suppress the super-divergent term. This is the motivation beyond the reduced parquet equations. 
%%
%
% 
%The major problem with the parquet equations, even in their reduced form, is that they are non-linear integral equations. It is hence impossible to solve them in other than a numerical way. We then lose, however, the control of the critical behavior of the Bethe-Salpeter equation, which is undesirable. Moreover, the numerical solution cannot be analytically continued from Matsubara to real frequencies. The critical region of the singularity in the Bethe-Salpeter equation is where the parquet approach and its two-particle self-consistency are irreplaceable. 

\subsection{Even and odd parts of the  self-energy}
\label{sec:Self-energy}

The reduced parquet equations determine the irreducible and reducible vertices in the scattering channel with a singularity in the respective Bethe-Salpeter equation. The full two-particle vertex  is then given as
\begin{multline}\label{eq:Gamma-sum}
\Gamma_{\sigma\bar{\sigma}}(\veck,i\omega_{n}, \veck',i\omega_{n'}; \vecq,i\nu_{m}) 
= \Lambda_{\sigma\bar{\sigma}}(\veck,i\omega_{n},\veck'  + \vecq,  i\omega_{ n' + m}) 
\\
+\ K_{\sigma\bar{\sigma}}(\veck,i\omega_{n},\veck',i\omega_{n'};\vecq , i\nu_{m })  
\,.
\end{multline}

We use the convention with $k,k'$ being the incoming and outgoing energy-momentum  of the electron and $q$ is the transfer energy-momentum between the electron and the hole for the repulsive interaction studied here. 

The one-particle propagators in the parquet equations are treated as input. In conserving theories the one- and two-particle Green functions are, however, related. The problem of correlated electron systems is that there are two ways to match consistently the one-particle self-energy with the two-particle vertex. One way is the Ward identity and the other is the dynamical Schwinger-Dyson equation \cite{Baym:1962aa}. We recently demonstrated that no approximate solution can obey both relations exactly \cite{Janis:2017aa}. Neither of the two relations may, however, be disregarded. The former is needed for thermodynamic consistency  and making the approximation conserving. The latter comes from the microscopic quantum dynamics. We cannot guarantee both relations with a single vertex and a single self-energy. We must either use a single self-energy and two two-particle vertices or vice versa. Ambiguity in the two-particle vertices leads to ambiguous criticality, thermodynamic inconsistencies and inability to continue the approximations beyond the critical point in the Bethe-Salpeter equation to the ordered phase. We are hence forced to use just a single two-particle vertex and introduce two self-energies to keep the approximate theories free of inconsistencies.    

\begin{figure}
\includegraphics[width=6cm]{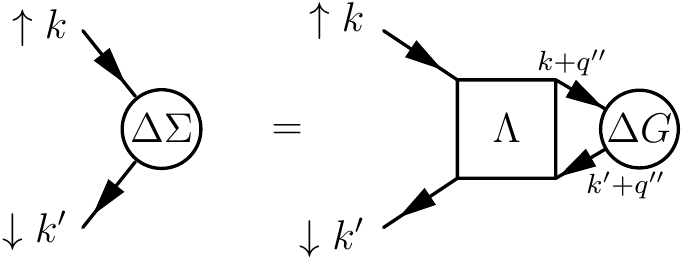}
\caption{Graphical representation of the odd part of the self-energy calculated from the normal part of irreducible vertex in the electron-hole channel, Eq.~\eqref{eq:SigmaT-general}. The odd part of the self-energy is anomalous in that it causes spin-flip. It is an order parameter that is non-zero only in the ordered phase.\label{fig:DeltaSigma} }
\end{figure}

 The two self-energies in approximate treatments can be introduced as even and odd parts of a single self-energy. The symmetry is set by the field controlling the critical fluctuations, conjugate to the order parameter. It is the longitudinal magnetic field in the present construction. The odd, or anomalous part of the self-energy will be determined via the Ward identity from the normal part, having the even symmetry, of the irreducible vertex $\Lambda_{\uparrow\downarrow}$ of the singular Bethe-Salpeter equation and the odd/anomalous part of the one-electron propagator. The Ward identity is, however, a functional differential equation that cannot be resolved for the self-energy exactly from the given two-particle vertex. To reach a qualitative thermodynamic consistency, it is sufficient to solve the Ward identity only linearly with respect to the symmetry-breaking field. The odd part or the thermodynamic self-energy is then defined as 
\begin{multline}\label{eq:SigmaT-general}
\Delta\Sigma(\veck,i\omega_{n}) =  - \frac 1N\sum_{\veck''}\frac 1{\beta}\sum_{\omega_{l}}
\\
\Lambda(\veck,i\omega_{n};\veck'', i\omega_{ l}) \Delta G(\veck'',i\omega_{l})\, ,
\end{multline}
where the normal part of the irreducible vertex with even symmetry with respect to the magnetic field is  $\Lambda(\veck,i\omega_{n},\veck'', i\omega_{l}) = \left[\Lambda_{\uparrow\downarrow}(\veck,i\omega_{n},\veck'', i\omega_{l}) + \Lambda_{\downarrow\uparrow}(\veck,i\omega_{n},\veck'', i\omega_{l})\right]/2$. The odd part of the one-electron propagator is  $\Delta G(\veck'',i\omega_{l}) = \left[G_{\uparrow}(\veck'',i\omega_{l}) - G_{\downarrow}(\veck'',i\omega_{l})\right]/2$. A diagrammatic representation of the Ward identity is presented in Fig.~\ref{fig:DeltaSigma}. It is evident that the anomalous self-energy is connected with the order parameter and vanishes in the spin-symmetric (paramagnetic) state, unlike the thermodynamic self-energy used in Refs.~\cite{Janis:2017aa,Janis:2017ab} 

\begin{figure}
\includegraphics[width=8.5cm]{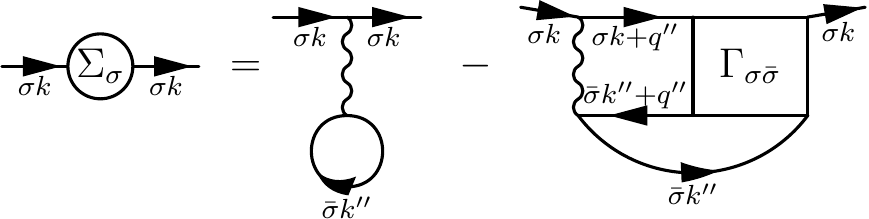}

\caption{Schwinger-Dyson equation from which the normal part with even symmetry with respect to the symmetry-breaking field is calculated, Eq.~\eqref{eq:SDE-spinpolarized}. The full vertex is the sum of the reducible and irreducible vertex from the electron-hole channel, $\Gamma_{\sigma\bar{\sigma}} = \Lambda_{\sigma\bar{\sigma}} + K_{\sigma\bar{\sigma}}$. \label{fig:SDE-figure}}
\end{figure}

The normal part of the  full self-energy  has even symmetry with respect to the magnetic field and will be determined from the Schwinger-Dyson equation for the spin-dependent self-energy
%
%\begin{widetext}
\begin{multline}\label{eq:SDE-spinpolarized}
\Sigma_{\sigma}(\veck,i\omega_{n}) = \frac U2 \left(n  - \sigma m\right) 
\\
-  \frac U{N^{2}}\sum_{\veck'',\vecq}\frac 1{\beta^{2}}\sum_{\omega_{l},\nu_{m}} G_{\sigma}(\veck'',i\omega_{l}) G_{\bar{\sigma}} (\veck'' + \vecq,i\omega_{l} + i\nu_{m})
 \\
 \times \Gamma_{\sigma\bar{\sigma}}(\veck'',i\omega_{l},\veck,i\omega_{n};\vecq, i\nu_{m}) 
 G_{\bar{\sigma}} (\veck + \vecq,i\omega_{n} + i \nu_{m})\,, 
\end{multline}
%\end{widetext}
where the total particle density $n$ and magnetization $m$ read 
\begin{align}\label{eq:n-full}
n &=  \frac 1{\beta N}\sum_{\veck,\sigma}\sum_{\omega_{n} } e^{i\omega_{n}0^{+}}G_{\sigma}(\veck, i\omega_{n}) \,,
\\
\label{eq:m-thermo}
m &= \frac 1{\beta N}\sum_{\veck,\sigma}\sum_{\omega_{n} } e^{i\omega_{n}0^{+}}\sigma G_{\sigma}(\veck, i\omega_{n}) \,.
\end{align}
 The full two-particle vertex is defined in Eq.~\eqref{eq:Gamma-sum} and the Schwinger-Dyson equation is diagrammatically represented in Fig.~\ref{fig:SDE-figure}.

This spin-polarized self-energy contains also odd powers of the magnetic field. The odd part of the self-energy is already determined by the Ward identity and the Schwinger-Dyson equation should  relate only the normal part of the self-energy with the two-particle vertex. The spin-dependent self-energy from Eq.~\eqref{eq:SDE-spinpolarized} must then be symmetrized to acquire the desired even symmetry with respect to the magnetic field. The normal part of the dynamical (spectral) self-energy  is
\begin{equation}\label{eq:Signma-Spectral}
\Sigma(\veck,\omega_{+}) = \frac 12\left[\Sigma_{\uparrow}(\veck,i\omega_{n}) + \Sigma_{\downarrow}(\veck,i\omega_{n}) \right]\,.
\end{equation}  
The full self energy in the one-particle propagator is a sum of  the anomalous self-energy, Eq.~\eqref{eq:SigmaT-general}, and the spectral self-energy, Eq.~\eqref{eq:Signma-Spectral}. The full one-electron propagator is 
\begin{multline}\label{eq:1P-Propagator}
G_{\sigma}(\veck,\omega)
\\
= \frac 1{\omega + \mu  - \epsilon(\veck)  + \sigma\left[ h -  \Delta \Sigma(\veck,\omega)\right] - \Sigma(\veck,\omega) }\,,
\end{multline}
where $\mu$ is the chemical potential, and $\epsilon(\veck)$ is the dispersion relation. 

Equations~\eqref{eq:SigmaT-general}-\eqref{eq:1P-Propagator} define a theory with the full one-particle self-consistency. The only difference from the Baym-Kadanoff approach is the decomposition of the self-energy into its normal and anomalous parts,  with even and odd symmetry with respect to the magnetic (symmetry-breaking) field determined from different exact equations. Should the Ward identity be compatible with the Schwinger-Dyson equation, we would recover the solution derived within the standard Baym and Kadanoff approach.   

It appears, as we showed in earlier publications and will demonstrate later also in this paper, that the full one-particle self-consistency does not necessarily lead to approximations with the best and most reliable results. It may be convenient to relax the one-particle self-consistency and to replace the normal part of the self-energy in the propagators used in the Bethe-Salpeter and Schwinger-Dyson equation by a simpler one, $\Sigma_{0}(\veck,\omega)$. It can be selected to optimize the approximate solution. If done so, we call the approximate one-electron Green function a thermodynamic propagator and denote it $G^{T}$. The same propagator is then used also in the equation determining the anomalous self-energy in the Ward identity, Eq.~\eqref{eq:SigmaT-general}.  We showed earlier that the simplest approximation best suited for the Kondo asymptotics in  SIAM is the Hartree approximation with $\Sigma_{0}(\veck,\omega)=Un/2$ \cite{Janis:2017aa}.

%\subsection{Critical region of the two-particle vertex}
%\label{sec:2Pcriticality}

\section{Mean-field approximation for the reducible vertex function}

\subsection{Mean-field-like decoupling of frequency convolutions}
\label{sec:MF-general}

The quantum character of many-body phenomena is manifested via the frequency dependence of the fundamental functions. Pure quantum critical behavior, free of spatial fluctuations, can be observed in the strong-coupling limit of impurity models  with correlated electrons. We now resort to a mean-field treatment and suppress the spatial fluctuations in the vertex functions and resort to a local theory. We now apply the general theory to the Kondo behavior of the single-impurity Anderson model to produce an impurity solver for systems with strongly correlated electrons. We keep the full frequency dependence.

The irreducible vertex from Eq.~\eqref{Lambda-Matsubara} in the local approximation is
\begin{subequations}\label{eq:Lambda-local}
\begin{equation}\label{eq:Lambda-dynamic}
\Lambda_{\sigma\bar{\sigma}}(i\omega_{n}, i\omega_{l}) = \frac U{1 + \mathcal{S}_{\sigma\bar{\sigma}}(i\omega_{l}, i\omega_{n})}\,,
\end{equation}   
with 
%\begin{widetext} 
\begin{multline}\label{eq:S-matsubara}
\mathcal{S}_{\sigma\bar{\sigma}}(i\omega_{l}, i\omega_{n})
= \frac 1\beta\sum_{\nu_{m}}K_{\sigma\bar{\sigma}}(i\omega_{l},i\omega_{n};-i\nu_{m}) 
\\
\times G_{\sigma }(i\omega_{n + m}) G_{\bar{\sigma}} (i\omega_{l - m}) \,.
\end{multline}
\end{subequations}
We recall the notation $\bar{\sigma} = -\sigma$. The equation for the reducible vertex reduces to an integral (matrix) equation in Matsubara frequencies
%\begin{widetext}
\begin{multline}\label{eq:K-Matsubara}
 \frac 1\beta\sum_{\omega_{l}}\left[\beta \delta_{n,l} + \Lambda_{\sigma\bar{\sigma}}(i\omega_{n},  i\nu_{m} + i\omega_{l})  G_{\bar{\sigma}} (i\nu_{m} + i\omega_{l})
 \right. \\ \left.
 G_{\sigma}(i\omega_{l}) \right]K_{\sigma\bar{\sigma}}(i\omega_{l},i\omega_{n'};i\nu_{m}) 
 =
  - \frac 1\beta\sum_{\omega_{l}} \Lambda_{\sigma\bar{\sigma}}(i\omega_{n}, i\omega_{l})
  \\
  \times  G_{\bar{\sigma}} (i\omega_{l}) G_{\sigma}(i\omega_{l})
\Lambda_{\sigma\bar{\sigma}}(i\omega_{l},,i\omega_{n'})  \,.
\end{multline}

Although the fermionic variables at low temperatures are relevant only near the Fermi energy, $\omega\approx 0$,  the integrals (sums) over the fermionic variables must be appropriately taken into account. We split the sum over Matsubara frequencies in sums over  positive and negative frequencies. We use the following notation
\be
\left\langle X(i\omega_{l})\right\rangle_{l} = \frac 1\beta\sum_{\omega_{l}> 0} X(i\omega_{l}) \,.
\ee

We cannot solve explicitly integral equation~\eqref{eq:K-Matsubara} but we can resort to a decoupling of the frequency convolutions in the spirit of a mean-value theorem for integrals of products of two positive functions. Fluctuations in the fermionic Matsubara frequencies are not relevant in the critical region of the singularity of the two-particle vertex. The fermionic frequencies are nevertheless important for keeping the approximation reliable also away from the critical region, in particular at non-zero temperatures.   We leave only one of the two-particle vertices $X$ and $Y$ in the convolution of type $\langle XGGY\rangle$ dynamic, frequency dependent, while the frequency dependence of the other vertex will be replaced by a single  mean value in the decoupling. It is reasonable to assume that the relevant values of the fermionic frequencies at low temperatures are only those from the vicinity of the Fermi energy.  We hence choose the mean value to be the limit to the Fermi energy from above for the sum over positive frequencies and from below for the sum over negative ones. The decoupling scheme of  the convolutions of fermionic Matsubara frequencies in the critical region of the low-temperature divergency of the reducible vertex then is 
%\begin{widetext}
\begin{multline} \label{eq:Convolutions-dynamical}
\sum_{\alpha=\pm 1}\left\langle X(i\omega_{n},i \omega_{\alpha l + m}) G(i\omega_{\alpha l + m}) G(i\omega_{\alpha l})Y(i\omega_{\alpha l}, i\omega_{n'})\right\rangle_{l} 
\\
\to \sum_{\alpha=\pm 1} \left\{X(i\omega_{n},0_{\alpha}) \left\langle  G(i\omega_{\alpha l }) G(i\omega_{\alpha l - m})Y(i\omega_{\alpha l - m}, i\omega_{n'})\right\rangle_{l} 
\right. \\ \left. 
+ \left\langle X(i\omega_{n},i \omega_{\alpha l + m}) G(i\omega_{\alpha l + m}) G(i\omega_{\alpha l}\right\rangle_{l})Y(0_{\alpha}, i\omega_{n'})
\right. \\ \left. 
 - X(i\omega_{n},i 0_{\alpha})\left\langle  G(i\omega_{\alpha l + m}) G(i\omega_{\alpha l})\right\rangle_{l}Y(0_{\alpha}, i\omega_{n'}) \right\} \,,
\end{multline}
%\end{widetext}
where we denoted $0_{\alpha} = i\eta\ \mathrm{sign}(\alpha)$ and $\eta > 0$ is infinitesimally small. This is a natural mean-field-like decoupling neglecting quadratic fluctuations beyond the mean value in averaging of products of operators. The decoupling of convolutions in Eq.~\eqref{eq:Convolutions-dynamical} holds for functions analytic in the upper and lower complex planes. That is what we assume about the Green and vertex functions in Eq.~\eqref{eq:K-Matsubara}. It is confirmed by explicit analytic continuation of sums in Eq.~\eqref{eq:K-Matsubara} to contour integrals resulting in an analytic expression for the reducible vertices  $\Lambda_{\sigma\bar{\sigma}}(z, z') $ and $K_{\sigma\bar{\sigma}}(z, z';\zeta)$ for $z,z',\zeta \in \mathbb{C}$. 

We can solve Eq.~\eqref{eq:K-Matsubara} explicitly with this decoupling of the frequency convolutions. This approximate solution generates a rich analytic structure of the vertex functions with cuts along the real axis of the fermionic frequencies as well as along positive and negative diagonals in the plane of complex frequencies. The calculations are, however, rather lengthy and we leave the application of the dynamical decoupling from Eq.~\eqref{eq:Convolutions-dynamical} to a separate publication. Here we resort to a simpler static decoupling of convolutions of fermionic frequencies  
\begin{multline} \label{eq:Convolutions-static}
\left\langle X(i\omega_{n},i \omega_{\alpha l + m}) G(i\omega_{\alpha l + m}) G(i\omega_{\alpha l})Y(i\omega_{\alpha l}, i\omega_{n'})\right\rangle_{l}
\\
 \to  X(i\omega_{n},i 0_{\alpha})\left\langle  G(i\omega_{\alpha l + m}) G(i\omega_{\alpha l})\right\rangle_{l}Y(0_{\alpha}, i\omega_{n'}) \,.
\end{multline}
The static decoupling is sufficient to gain a first qualitative picture of the strong-coupling regime. It offers a generalization of the static approximation from Ref.~\cite{Janis:2017aa} to the spin-polarized state and to non-zero temperatures.

\subsection{Static mean-field approximation}

We apply the general theory with the static decoupling of convolutions of fermionic Matsubara frequencies, Eq.~\eqref{eq:Convolutions-static},  to the strong-coupling limit of SIAM where the Kondo effect is observed at half filling and zero temperature.  The static decoupling with the relevant values of the irreducible vertex only from the Fermi energy simplifies the solution of the reduced parquet equations significantly without missing the strong-coupling limit. We then have $\Lambda_{\uparrow\downarrow}(\omega_{\sigma}, \omega'_{\tau}) \to \Lambda_{\uparrow\downarrow}(0_{\sigma},0_{\tau})$.   Consequently we reduce the reducible vertex to $K_{\uparrow\downarrow}(\omega' + i\tau\eta',\omega + i\sigma \eta;\Omega_{\rho})  \to \mathcal{K}_{\uparrow\downarrow}(i\tau 0^{+}, i\sigma 0^{+};\Omega_{\rho}) = K_{\uparrow\downarrow}(\tau, \sigma)/D_{\uparrow\downarrow}(\Omega_{\rho})$, where $D_{\uparrow\downarrow}(\Omega_{\rho})$ is the determinant of the matrix of the kernel of the equation for the reducible vertex $\mathcal{K}_{\uparrow\downarrow}(i\tau 0^{+}, i\sigma 0^{+};\Omega_{\rho})$ with $\Omega\to 0$. 

 One can solve the equations to determine $\Lambda_{\uparrow\downarrow}(0_{\sigma},0_{\tau})$ and $K_{\uparrow\downarrow}(\tau, \sigma)$. The reduction of the dependence of the vertex functions on the fermionic frequencies to their values at the Fermi energy is justified and works reliably in the quantum critical region of the singularity of the two-particle vertex, that is at zero temperature. Both vertex functions of fermionic frequencies are continuous across the Fermi energy at zero temperature, that is $\Lambda_{\uparrow\downarrow}(\sigma, \tau) =\Lambda_{\uparrow\downarrow}$ and $K_{\uparrow\downarrow}(\tau, \sigma) = K_{\uparrow\downarrow}$. The two values   $\Lambda_{\uparrow\downarrow}(+, +)$ and $\Lambda_{\uparrow\downarrow}(+,-)$ split at non-zero temperatures where the values at the Fermi energy play gradually less and less dominant role.   If we want to extrapolate consistently the low-temperature approximation also to  high temperatures far above the Kondo temperature, one has to return to the full dynamical decoupling of convolutions of fermionic frequencies, Eq.~\eqref{eq:Convolutions-dynamical} and keep the full dynamics of the irreducible vertex $\Lambda_{\uparrow\downarrow}(\omega_{\sigma}, \omega'_{\tau}) $. A simpler option without leaving the static approximation is  to select a single value of the irreducible vertex at the Fermi energy that screens the bare interaction efficiently at all temperatures and is dominating at high temperatures. It is  $\Lambda_{\uparrow\downarrow}(0_{+},0_{-})$  that does the job and will be used to extend the zero-temperature static solution continuously to high temperatures \cite{Janis:2019ab}. 
 
 The single irreducible vertex reduces then to an effective interaction defined from an equation 
\begin{equation}\label{eq:Lambda-static}
\Lambda_{\uparrow\downarrow} \equiv \Lambda_{\uparrow\downarrow}(0_{+},0_{-}) 
 =  \frac U{1 + \mathcal{K}_{\uparrow\downarrow}X_{\uparrow\downarrow}}
\end{equation}
where
\begin{multline}
 X_{\uparrow\downarrow}  = \int_{-\infty}^{\infty}\frac{dx}{\pi} \left\{\frac{\Re\left[ G_{\uparrow}(x_{+})G_{\downarrow}(-x_{+})\right]}{\sinh(\beta x)} 
\right. \\ \left. 
\times \Im\left[ \frac 1{D_{\uparrow\downarrow}(-x_{+})}\right] - f(x) \Im\left[\frac {G_{\uparrow}(x_{+}) G_{\downarrow}(-x_{+})}{D_{\uparrow\downarrow}(-x_{+})} \right] \right\} \,.
\end{multline}
We denoted $x_{+} = x + i0^{+}$ and used an equality $f(x) + b(x) = 1/\sinh(\beta x)$. We straightforwardly analytically continued the sums over Matsubara frequencies to spectral integrals with the Fermi, $f(x) = 1/(e^{\beta x} + 1)$ and the Bose, $b(x) = 1/(e^{\beta x} - 1)$, distributions. 

The reduced parquet equations now have an explicit algebraic solution. The full, frequency dependent determinant $D_{\uparrow\downarrow}(\Omega_{+})$ reads as
%\begin{widetext}
\begin{multline}\label{eq:Det-Omega}
D_{\uparrow\downarrow}(\Omega_{+}) = 1 + \Lambda_{\uparrow\downarrow} \left[\left\langle G_{\downarrow}(x + \Omega_{+}) \Im G_{\uparrow}(x_{+})\right\rangle_{x} 
\right. \\ \left. 
+\ \left\langle G_{\uparrow}(x - \Omega_{+}) \Im G_{\downarrow}(x_{+})\right\rangle_{x}\right]   \,.
\end{multline}
%
%\end{widetext}
where we denoted 
\begin{multline}
\left\langle G_{s}(x + \omega) G_{s'}(x + \omega')\right\rangle_{x} = -\int_{-\infty}^{\infty} \frac{dx}{\pi} f(x) 
\\
\times G_{s}(x + \omega) G_{s'}(x + \omega') \,.
\end{multline}

The equation for the reducible vertex with frequencies near the Fermi energy $ \mathcal{K}_{\uparrow\downarrow}= \mathcal{K}_{\uparrow\downarrow}(0_{-},0_{+})$ reads as
%\begin{widetext}
%\begin{subequations}\label{eq:K-static}
\begin{equation}\label{eq:K-mp}
\mathcal{K}_{\uparrow\downarrow} = - \Lambda_{\uparrow\downarrow}^{2}\left\langle \Im \left[G_{\downarrow}(x_{+}) G_{\uparrow}(x_{+})\right]\right\rangle_{x}
\end{equation}
%\end{subequations}
%\end{widetext}

We denote $g_{\uparrow\downarrow}(+) = \left\langle \Im \left[G_{\downarrow}(x_{+}) G_{\uparrow}(x_{+})\right]\right\rangle_{x}$ and introduce a dimensionless Kondo scale as the zero value of function $D_{\uparrow\downarrow}(\Omega_{+})$ from Eq.~\eqref{eq:Det-Omega}, 
\begin{equation}\label{eq:Det-zero}
a= D_{\uparrow\downarrow}(0) = 1 +  \Lambda_{\uparrow\downarrow}g_{\uparrow\downarrow}(+) \,.  
\end{equation}
It measures the distance to the critical point $a=0$ and will affect the low-energy behavior in the strong-coupling limit.

We  rewrite Eq.~\eqref{eq:K-mp} by using Eq.~\eqref{eq:Lambda-static} to another, more suitable form  
\begin{equation}\label{eq:Lambda-m}
1 > 1 - \frac{U - \Lambda_{\uparrow\downarrow}}{X_{\uparrow\downarrow}\Lambda_{\uparrow\downarrow}^{2}} =  D_{\uparrow\downarrow}(0) > 0\,. 
\end{equation}
The right inequality guarantees stability of the solution and integrability and positivity of the integral $X_{\uparrow\downarrow}$ in the strong-coupling regime.  

Equations~\eqref{eq:Lambda-static}-\eqref{eq:K-mp} form a closed set of equations  determining self-consistently the values of the vertex functions $\Lambda_{\uparrow\downarrow}$ and $\mathcal{K}_{\uparrow\downarrow}$ at the Fermi energy. They can be solved numerically in a straightforward way via iterations. The bare interaction and the one-electron propagators are input to these equations. The latter contain the normal and anomalous parts of the self-energy. The anomalous self-energy renormalizing the effect of the magnetic field in the static approximation is
\begin{equation}
\Delta\Sigma = \frac{\Lambda}2 \left[\left\langle \Im G_{\downarrow}(x_{+})\right\rangle_{x}  - \left\langle \Im G_{\uparrow}(x_{+})\right\rangle_{x}\right]  \,.
\end{equation}    
We recall that $\Lambda = \left(\Lambda_{\uparrow\downarrow} + \Lambda_{\downarrow\uparrow}\right)/2$ is the normal part of the irreducible vertex. But due to the electron-hole symmetry $\Lambda_{\uparrow\downarrow} = \Lambda_{\downarrow\uparrow}$.

The spin-polarized self-energy from the Schwinger-Dyson equation is
%\begin{widetext}
\begin{multline}
\Sigma_{\uparrow}(\omega_{+}) = \frac U2 \left(n  - \sigma m\right)
\\
+\   U\int_{-\infty}^{\infty}\frac{dx}{\pi } \left\{f(x + \omega) \frac{\phi_{\uparrow\downarrow}( x _{-})}{D_{\uparrow\downarrow}(x_{-})}  \Im G_{\downarrow}(x_{+} + \omega)
\right. \\ \left.
 -\ b(x)G_{\downarrow}(x_{+} + \omega) \Im_{x}\left[\frac{\phi_{\uparrow\downarrow}( x _{+})}{D_{\uparrow\downarrow}(x_{+})} \right]  \right\}\,,
\end{multline}
where
\begin{multline} \label{eq:Lambda-phi}
\phi_{\uparrow\downarrow}(\Omega_{\pm}) 
\\
= - \Lambda_{\uparrow\downarrow} \int_{-\infty}^{\infty}\frac{dx}{\pi} f(x) \left[G_{\downarrow}(x + \Omega_{\pm})\Im  G_{\uparrow}(x_{+}) 
\right. \\ \left.
+\ G_{\uparrow}(x - \Omega_{\pm})\Im  G_{\downarrow}(x_{+}) \right]\,,
\end{multline}
is the electron-hole bubble multiplied by the effective interaction. 

The normal part of the dynamical self-energy then is $\Sigma(\omega_{+}) = \left[\Sigma_{\uparrow}(\omega_{+}) + \Sigma_{\downarrow}(\omega_{+}) \right]/2$ and the one-electron propagator with the full self-energy for SIAM is  
\begin{equation}\label{eq:1P-Propagator-SIAM}
G_{\sigma}(\omega)
= \frac 1{\omega + \mu + \sigma\left( h -  \Delta \Sigma\right) - \Sigma(\omega)  + i \Delta}\,,
\end{equation}
where $\Delta$ is the width of the local level attached to conducting leads and is set as the energy unit. 

The present construction allows for selecting the optimal degree of the one-electron self-consistency. It means that the propagators used in the perturbation expansion, in the Ward identity, the Bethe-Salpeter, and the Schwinger-Dyson equations, need not be the fully renormalized propagator from Eq.~\eqref{eq:1P-Propagator-SIAM}. We can use a thermodynamic propagator with an appropriately chosen normal self-energy $\Sigma_{0}(\omega)$ with which we control the degree of one-particle self-consistency 
\be
G_{\sigma}^{T}= \frac 1{\omega + \mu - \frac U2 n^{T} + \sigma(h - \Delta\Sigma) - \Sigma_{0}(\omega)  + i \Delta}
\ee
where the Hartree term is calculated with the thermodynamic propagator 
\be
n^{T} = - \sum_{\sigma}\int_{-\infty}^{\infty}\frac{dx}{\pi} f(x) \Im G_{\sigma}^{T}(x_{+}) \,.
\ee
It appears that to reach the best fit for the asymptotic form of the Kondo scale is $\Sigma_{0}(\omega) = 0$, which we use in the thermodynamic propagator $G^{T}_{\sigma}(\omega)$ in the following section.

\section{Strong-coupling limit at zero temperature of SIAM}

The critical behavior of SIAM emerges in the asymptotic limit to strong-coupling of the  spin- and charge-symmetric state at zero temperature. We now explicitly solve the equations within the static decoupling of the frequency convolutions at zero temperature where the irreducible vertex is continuous at the Fermi energy. We will analyze the spin polarized solution generally also away from half filling to demonstrate how the critical behavior fades away when moving from the spin and charge symmetric situation and leaving the strong-coupling regime.

\subsection{Analytic solution}

The irreducible vertex, the effective interaction at zero temperature is 
\begin{equation}\label{eq:Lambda-zeroT}
\Lambda_{\uparrow\downarrow} 
 =  \frac U{\displaystyle 1 - K_{\uparrow\downarrow}\int_{-\infty}^{0}\frac {dx}{\pi}\Im\left[\frac {G_{\uparrow}(x_{+}) G_{\downarrow}(-x_{+})}{D_{\uparrow\downarrow}(-x_{+})} \right] }
\end{equation}
with the determinant $D_{\uparrow\downarrow}(x_{+})$ and $K_{\uparrow\downarrow}$ defined in Eqs.~\eqref{eq:Det-Omega} and~\eqref{eq:K-mp}. 

We separately represent the imaginary and real parts of the dynamical self-energy. The imaginary part has the following representation
\begin{subequations}\label{eq:Sigma-sp-0}
\begin{multline}\label{eq:Sigma-sp-0-Im}
\Im\Sigma_{\uparrow}(\omega_{+}) =  U \int_{-|\omega|}^{|\omega|}\frac{dx}{\pi}\Im G_{\downarrow}(x + \omega_{+}) \Im\left[\frac{\phi_{\uparrow\downarrow}(x_{+})}{D_{\uparrow\downarrow}(x_{+})}\right]
 \\ 
 \times \left[\theta(\omega) \theta(-x) - \theta(-\omega) \theta(x) \right]\,,
\end{multline}
while the real part is
\begin{multline}\label{eq:Sigma-sp-0-Re}
\Re\Sigma_{\uparrow}(\omega_{+}) =  U \int_{-\infty}^{-\omega}\frac{dx}{\pi} \Re\left[\frac{\phi_{\uparrow\downarrow}(x_{+}) }{D_{\uparrow\downarrow}(x_{+})}\right] \Im G_{\downarrow}(x + \omega_{+})
 \\ 
 +\  U \int_{-\infty}^{0}\frac{dx}{\pi} \Im\left[\frac{\phi_{\uparrow\downarrow}(x_{+})}{D_{\uparrow\downarrow}(x_{+})}\right] \Re G_{\downarrow}(x + \omega_{+}) \,.
\end{multline}
\end{subequations}
Function $\phi_{\uparrow\downarrow}(x_{+})$ was defined in Eq.~\eqref{eq:Lambda-phi}. 

The thermodynamic propagator can be represented via an effective chemical potential and an effective magnetic field
 \begin{equation}\label{eq:GT}
G_{\sigma}^{T}(\omega_{+}) = \frac {1}{\omega + \bar{\mu}  + \sigma\bar{h} + i\Delta} \,,
\end{equation}
that are derived from the bare chemical potential and magnetic field together with the  thermodynamic particle density and magnetization
\begin{align}\label{eq:bar-mu}
\bar{\mu} &= \mu - \frac U2 n^{T} \,,\\
\bar{h} &= h + \frac \Lambda2 m^{T} \,. 
\end{align}
The unrenormalized thermodynamic propagator allows for analytic representations of thermodynamic quantities. The thermodynamic charge density and magnetization are 
\begin{align}\label{eq:nT-def}
n^{T} &= 1 + \frac 1\pi\left[\arctan(\bar{\mu} + \bar{h}) + \arctan(\bar{\mu} - \bar{h})\right]  \,,\\
m^{T} &= \frac 1\pi\left[\arctan(\bar{\mu} + \bar{h}) - \arctan(\bar{\mu} - \bar{h})\right]  \,. 
\end{align}

The zero-field susceptibility from the thermodynamic propagator is
\begin{equation}\label{eq:chi-T}
\chi^{T}= \left.\frac{d m^{T}}{d h}\right|_{h=0} = \frac {2}{1 + \Lambda \phi(0)}\int_{-\infty}^{0} \frac{dx}{\pi} \Im\left[G^{T}(x_{+})^{2} \right] \,. 
\end{equation}

The full physical spin-dependent propagator is
\begin{equation}\label{eq:G-renorm}
G_{\sigma}(\omega_{+}) 
= \frac 1{\omega + \mu - \frac U2 n + \sigma \bar{h} - \Sigma(\omega_{+}) + i \Delta} \,.
\end{equation}
 The physical particle density and magnetization are 
\begin{align}\label{eq:n-def}
n &= -\sum_{\sigma}\int_{-\infty}^{0}\frac{dx}{\pi} \Im G_{\sigma}(x_{+}) \,, \\
m &= -\sum_{\sigma}\sigma\int_{-\infty}^{0}\frac{dx}{\pi} \Im G_{\sigma}(x_{+}) \,, 
\end{align}
respectively and they differ slightly from the thermodynamic ones. 
 The zero-field physical susceptibility from the full propagator is 
\begin{equation}
\chi = \left.\frac{d m}{d h}\right|_{h=0} = \left(2 + \Lambda \chi^{T} \right)\int_{-\infty}^{0} \frac{dx}{\pi} \Im\left[G(x_{+})^{2} \right].
\end{equation}

Only small frequencies in $D_{\uparrow\downarrow}(\omega_{+})$ are relevant in the strong-coupling limit. We can then replace the full frequency dependence in $D_{\uparrow\downarrow}(\omega_{+})$ by a low-frequency expansion, keeping only the term linear in frequency. That is, 
\begin{equation}\label{eq:D-low}
D_{\uparrow\downarrow}(\omega_{+}) \doteq a - iD'_{\uparrow\downarrow}\omega = a - i(D_{R} + i D_{I})\omega \,.
 \end{equation}
The linear coefficient reads as
\begin{multline}
D'_{\uparrow\downarrow}  = \left\langle \partial G_{\downarrow}(x_{+})\Im G_{\uparrow}(x_{+}) - \partial G_{\uparrow}(x_{-})
\right. \\ \left.
\times \Im G_{\downarrow}(x_{+})\right\rangle_{x} \Lambda_{\uparrow\downarrow} = \left[ i\left\langle \Im G_{\uparrow}^{T}(x_{+}) \partial \Re G_{\downarrow}(x_{+}) 
\right.\right. \\ \left.\left.
- \Im G_{\downarrow}(x_{+})\partial \Re G_{\uparrow}(x_{-})\right\rangle_{x}  +\pi\rho_{\uparrow}\rho_{\downarrow}\right]\Lambda_{\uparrow\downarrow}  \,, 
\end{multline}
where $\rho_{\sigma}= -\Im G_{\sigma}(0_{+})/\pi$ is the density of states at the Fermi energy.

The theory with the full one-particle self-consistency and the full propagator in all equations can be solved only numerically at intermediate and not too strong electron interaction. The theory with the thermodynamic propagator, on the other hand, allows for an explicit analytic representation also in the Kondo limit. It is defined when $a= 1 + \Lambda g_{\uparrow\downarrow}(0) \ll 1$. Using the thermodynamic propagator from Eq.~\eqref{eq:GT} and the low-frequency asymptotics of the determinant from Eq.~\eqref{eq:D-low} we can explicitly evaluate the frequency integrals. 
 
 The asymptotic form of the effective interaction in strong coupling within the low-frequency asymptotics, Eq.~\eqref{eq:D-low} is 
 \begin{multline}\label{eq:Lambda-asymptotic}
 \Lambda = \frac{U}{1 + \displaystyle{\frac{\Lambda D_{0}|\ln a|}{\pi\left[D_{R}^{2} + D_{I}^{2}\right]}}\left\{\Re\left[G_{\uparrow}G_{\downarrow}^{*} \right] D_{R} - \Im\left[G_{\uparrow}G_{\downarrow}^{*} \right] D_{I} \right\}}
 \\
 = \frac{U}{1 + \displaystyle{\frac{\Lambda D_{0}|\ln a|\left[(\bar{\mu}^{2} - \bar{h}^{2} + 1) D_{R} - 2 \bar{h} D_{I}\right]}{\pi \left[D_{R}^{2} + D_{I}^{2}\right]\left[(\bar{\mu}^{2} - \bar{h}^{2} + 1)^{2} + 4 \bar{h}^{2}\right]}}} \,,
  \end{multline} 
where $G_{\sigma}= G_{\sigma}(0_{+})$ and   
\begin{multline}
 D_{0}  =  -\phi_{\uparrow\downarrow}(0) = \int_{-\infty}^{0}\frac{dx}{\pi}\Im\left[G_{\uparrow}(x_{+})G_{\downarrow}(x_{+})\right] = \frac 1{2\pi \bar{h}}
 \\ 
 \times\left[ \arctan \frac{2\bar{h}}{1 + \bar{\mu}^{2} - \bar{h}^{2}} 
 %\right. \\ \left.
 + \pi \theta(\bar{h}^{2} - \bar{\mu}^{2} - 1) \mathrm{sign}(\bar{h})\right] \,,
 \end{multline}
 \begin{multline}
 D_{R}  =\pi \rho_{\uparrow}\rho_{\downarrow} = \frac 1{\pi}\frac{1}{(\bar{\mu} + \bar{h})^{2} + 1}
\frac{1}{(\bar{\mu} - \bar{h})^{2} + 1}\,,
 \end{multline}
\begin{multline}
 D_{I} =  -\int_{-\infty}^{0}\frac{dx}{\pi} \left[\Im G_{\uparrow}(x_{+}) \partial \Re G_{\downarrow}(x_{+}) 
 \right. \\ \left.
 - \Im G_{\downarrow}(x_{+})\partial \Re G_{\uparrow}(x_{-})\right]
  \\ 
= \frac 1{2\pi \bar{h}} \left\{\frac{1 + \bar{\mu}^{2} - \bar{h}^{2}}{(\bar{\mu}^{2} - \bar{h}^{2})^{2} + 2(\bar{\mu}^{2} + \bar{h}^{2}) + 1} 
 \right.\\ \left.
 - \frac 1{2\bar{h}} \left[\arctan \frac{2\bar{h}}{1 + \bar{\mu}^{2} - \bar{h}^{2}} + \pi \theta(\bar{h}^{2} - \bar{\mu}^{2} - 1) \mathrm{sign}(\bar{h})\right]\right\} \,.
 \end{multline} 
 The first equality holds always generally for arbitrarily renormalized one-electron propagator while the second one only for the thermodynamic propagator $G^{T}(\omega_{+})$ from Eq.~\eqref{eq:GT}. Since the solution is a local Fermi liquid where $\Sigma(0) = 0$, the actual degree of the one-particle renormalization does not play a  role in the strong-coupling asymptotics.  
 Notice that the imaginary part of the linear coefficient of the determinant $D_{\uparrow\downarrow}(\omega_{+})$ is non-zero only in the spin-polarized state. 
  
The asymptotic form of the imaginary part of the spin-polarized dynamical self-energy in strong coupling is
\begin{subequations}\label{eq:SE-asympt}
\begin{multline}
\Im\Sigma_{\sigma}(\omega_{+}) 
\\
= \frac{UD_{0}\Im G_{\bar{\sigma}} ^{T}(\omega_{+})}{\pi \left[D_{R}^{2} + D_{I}^{2}\right]} \left\{ D_{R}\ln \sqrt{1  + \frac{\omega^{2}\left[D_{R}^{2} 
+ D_{I}^{2}\right]}{\bar{a}^{2}} }  
\right. \\ \left.
+\ D_{I}\left[\arctan\frac{D_{R}\omega}{\bar{a} - D_{I}\omega} + \pi \theta(D_{I}\omega - \bar{a}) \mathrm{sign} (\omega)\right]\right\}\,
\end{multline}
with the corresponding real part obeying the causality condition 
\begin{multline}
\Re\Sigma_{\sigma}(\omega_{+}) 
\\
= - \frac{UD_{0}\Im G_{\downarrow}^{T}(\omega_{+})}{\pi \left[D_{R}^{2} + D_{I}^{2}\right]} \left\{D_{I}\ln \sqrt{\left(\bar{a} - D_{I}\omega \right)^{2} + D_{R}^{2}\omega^{2}} 
\right. \\ \left.
-\ D_{R}\left[\arctan\left( \frac{D_{R}\omega}{\bar{a} - D_{I}\omega}\right) + \pi \theta(D_{I}\omega - \bar{a}) \mathrm{sign}  (\omega)\right]\right\}
\\
+\ \frac{UD_{0}D_{R} |\ln \bar{a}|\Re G_{\downarrow}(\omega_{+})}{\pi \left[D_{R}^{2} + D_{I}^{2}\right]} \,, 
\end{multline}
\end{subequations}
where we denoted $\bar{a} = a/\Lambda$.

Finally, we can explicitly evaluate the Kondo scale  as a function of the effective chemical potential and the effective magnetic field when we  use the critical effective interaction $\Lambda = 1/D_{0}$. We then obtain  
\begin{multline}\label{eq:Kondo-log}
  -\ln a_{K} = \frac{\pi (UD_{0} - 1)\left[D_{R}^{2} + D_{I}^{2}\right]}{\Re\left[G_{\uparrow}G_{\downarrow}^{*}\right]D_{R} - \Im\left[G_{\uparrow}G_{\downarrow}^{*}\right]D_{I}}
  \\
  = \frac{\pi (UD_{0} - 1)\left[D_{R}^{2} + D_{I}^{2}\right]\left[(\bar{\mu}^{2} - \bar{h}^{2} + 1)^{2} + 4 \bar{h}^{2}\right]}{\left[(\bar{\mu}^{2} - \bar{h}^{2} + 1) D_{R} - 2 \bar{h} D_{I}\right]}\,.
  \end{multline}
This explicit representation can be used to determine the boundary of the strong-coupling region. It is set to the point where the logarithm of the asymptotic Kondo scale goes through zero, that is, $UD_{0}=1$. The Kondo regime is then defined for the interaction strength obeying $UD_{0}\gg 1$.

\subsection{Numerical results}

We used thermodynamic and full one-electron propagators  to compare the results with the exact ones for SIAM at zero temperature. Since the equation for the irreducible vertex, Eq.~\eqref{eq:Lambda-zeroT}, slightly differs  from the vertex used in Refs.~\cite{Janis:2017aa,Janis:2017ab} we also compare the two versions of the two-particle self-consistency.  

\begin{figure}
\includegraphics[width=7.5cm]{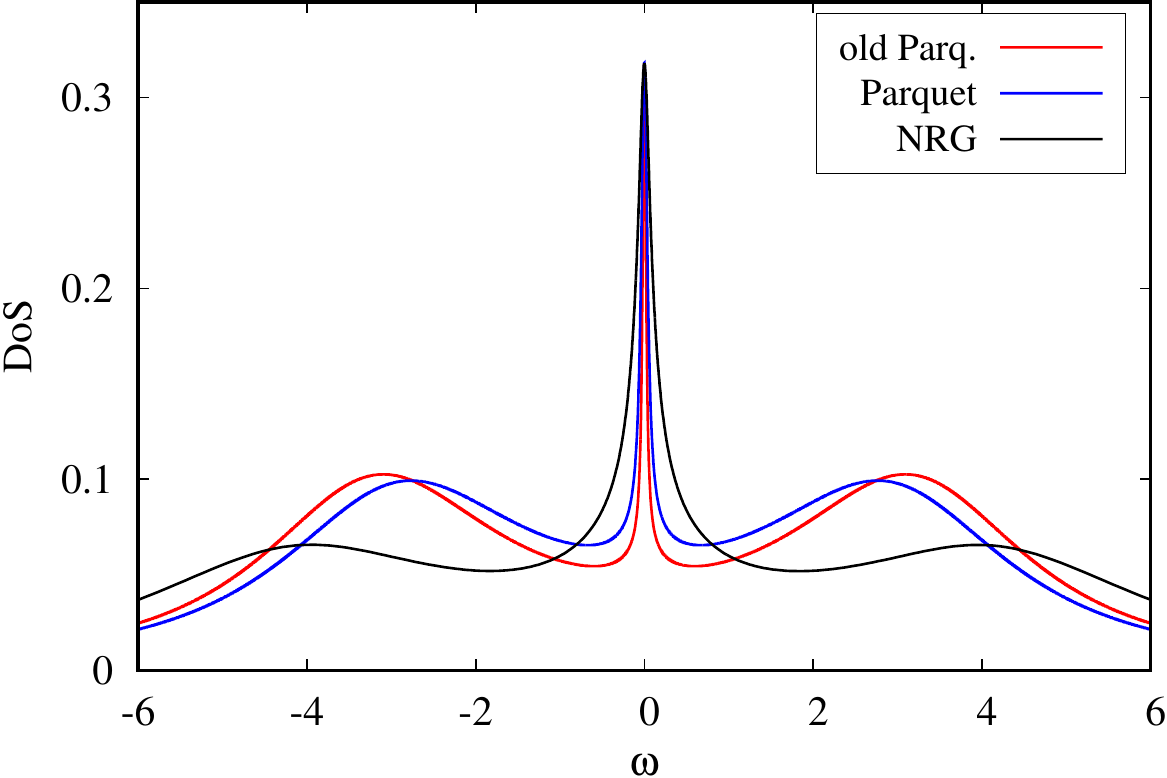}
\caption{ Spectral function at $U/\Delta =8$ at half filling calculated within the static approximation, Eq.~\eqref{eq:Lambda-zeroT} (Parquet), Ref.~\cite{Janis:2017aa} (old Parq.), and NRG in energy units $\Delta$. \label{fig:U8Sp}}
\end{figure}

The Kondo effect and the Kondo strong-coupling regime occur at half filling of the non-magnetic state. We compare in Fig.~\ref{fig:U8Sp} the present version of the static approximation with the thermodynamic propagator from Eq.~\eqref{eq:GT}  with that of Ref.~\cite{Janis:2017aa} and with the spectral function from NRG. There is  no much difference between the two versions of the static approximation. They both have the same enhancement of the satellite Hubbard bands with a narrower central Kondo-Suhl resonance compared to NRG.  

The NRG calculations were performed with the NRG Ljubljana code~\cite{NRGLjubljana}. A constant density of states of bandwidth $2D$ with $U/2D > 100$ was used. Spectral functions were obtained from the DM-NRG algorithm of Ref.~\cite{Hofstetter:2000aa}. We opted for not correcting the spectral energies via the so-called self-energy trick. All results were recalculated from the typical NRG units of $D$ into the units of $\Delta$ as used in this paper.  

\begin{figure}
\includegraphics[width=8.5cm]{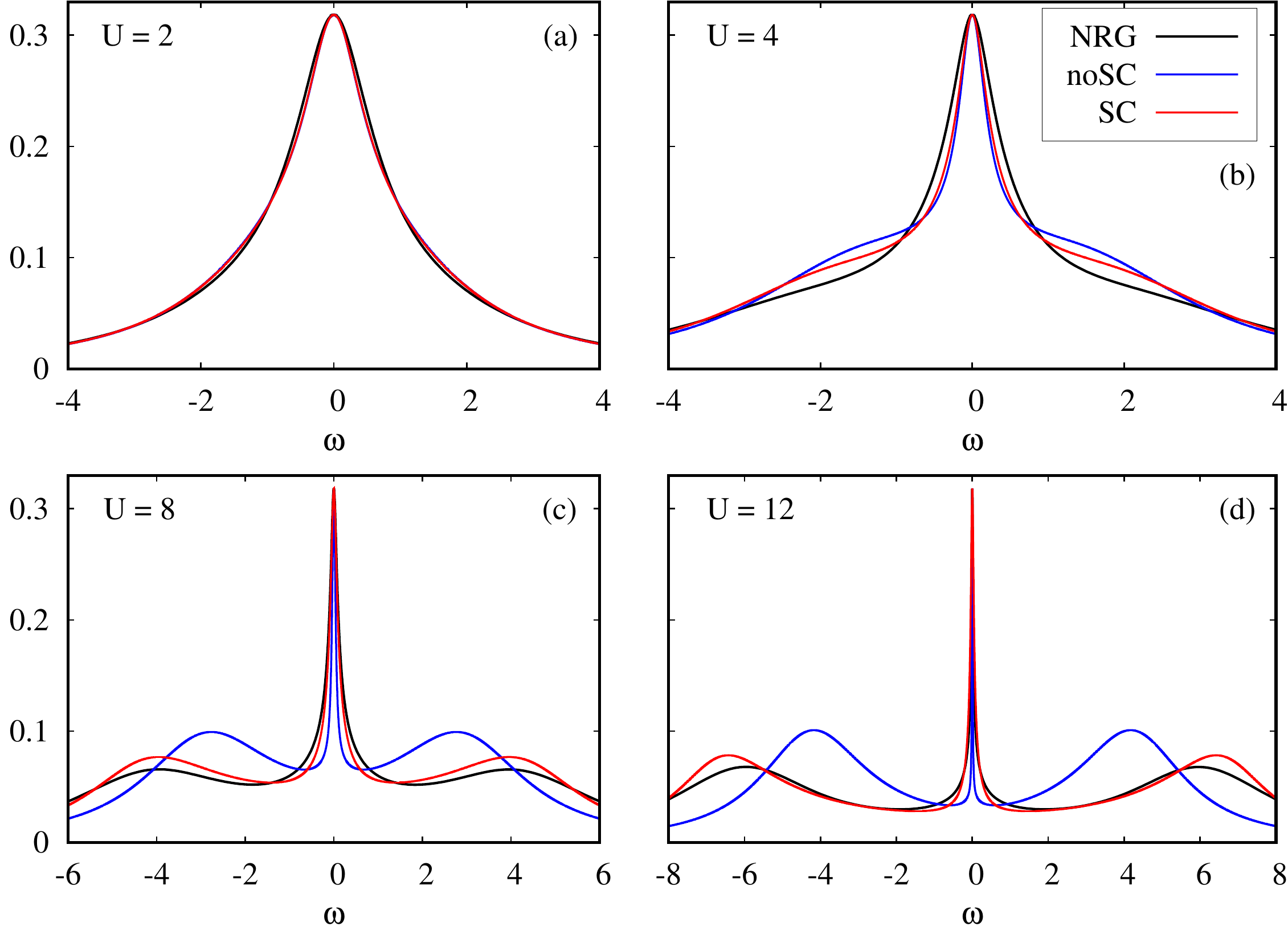}

\caption{Spectral function calculated within the static approximation with the thermodynamic propagator, Eq.~\eqref{eq:GT} (noSC), full propagator, Eq.~\eqref{eq:G-renorm} (SC) in the reduced parquet equations compared with NRG for several values of the interaction in energy units $\Delta$.  \label{fig:SpUvar}}
\end{figure}
\begin{figure}
\includegraphics[width=7.5cm]{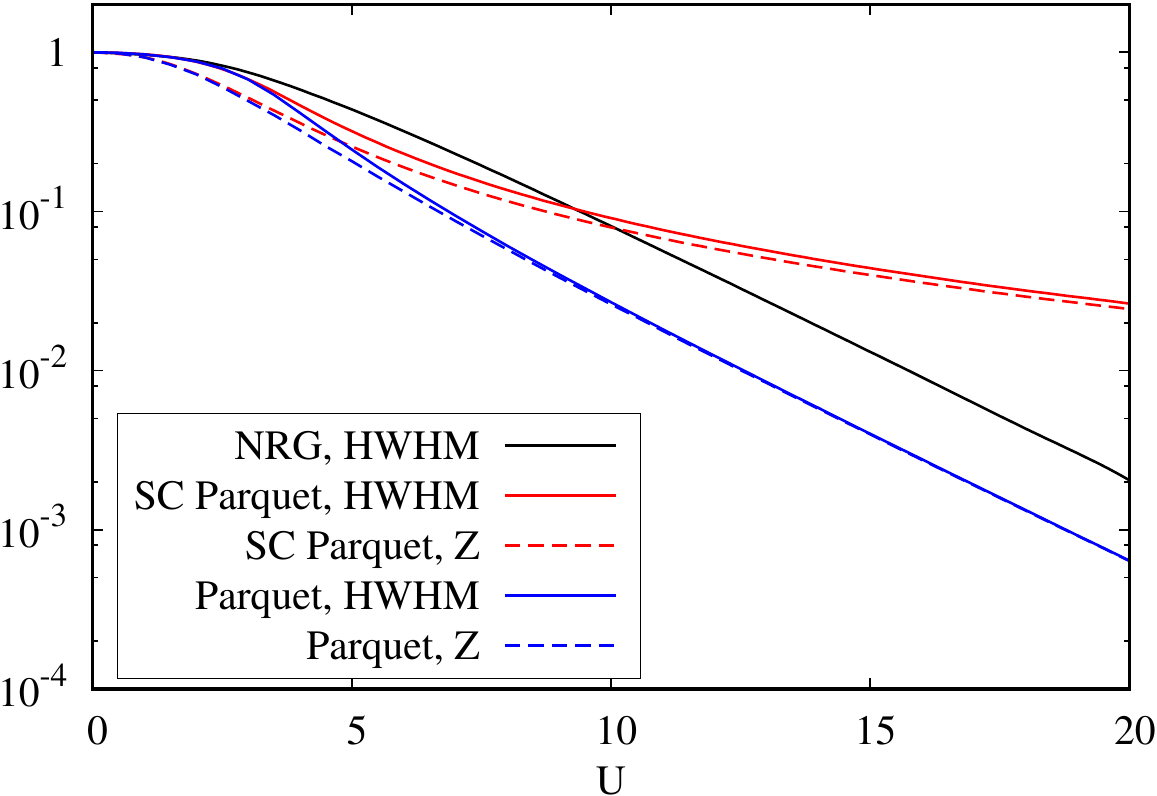}

\caption{Half width of the Kondo-Suhl  quasiparticle peak at half maximum (HWHM) and $Z$ factor calculated with one-particle self-consistency (SC  Parquet) and with the thermodynamic propagator (Parquet)  compared with the NRG result in energy units $\Delta$.  \label{fig:Kondo-symm}}
\end{figure}
One can improve upon the thermodynamic propagator and take into account the full one-particle self-consistency. We take the same full propagator in the parquet equations as well as in the Ward identity and the Schwinger-Dyson equation. We compared the two solutions for the spectral function of the non-magnetic state at half filling with the NRG result for weak and moderate interactions in Fig.~\ref{fig:SpUvar}. In weak coupling all three approaches give almost the same function. In intermediate coupling the fully self-consistent version delivers a better agreement with the NRG result for high frequencies and in positioning of the Hubbard satellite bands. The width of central quasiparticle peak is, however, missed in the strong-coupling regime of the self-consistent version, Fig.~\ref{fig:Kondo-symm}. The non-self-consistent version with the thermodynamic propagator from Eq.~\eqref{eq:GT} predicts correctly the linear dependence of the logarithm of the Kondo scale on the interaction strength defined as the half width at half maximum (HWHM) of the central peak while the exponent of the self-consistent solution is one third.  We plotted also another definition of the Kondo scale from factor $Z= (1 - d\Sigma/d\omega|_{\omega=0})^{-1}$ which shows the same strong-coupling asymptotics. The shift of the curve calculated from our static approximation with the thermodynamic propagator is caused by a difference in the non-universal pre-factor at the logarithm of the Kondo scale. It is $\pi/8$ in the Bethe-ansatz solution with the Lorentzian density of states \cite{Hewson:1993aa}  while it comes as $1/\pi$ from our parquet equations.   

\begin{figure}
\includegraphics[width=7.5cm]{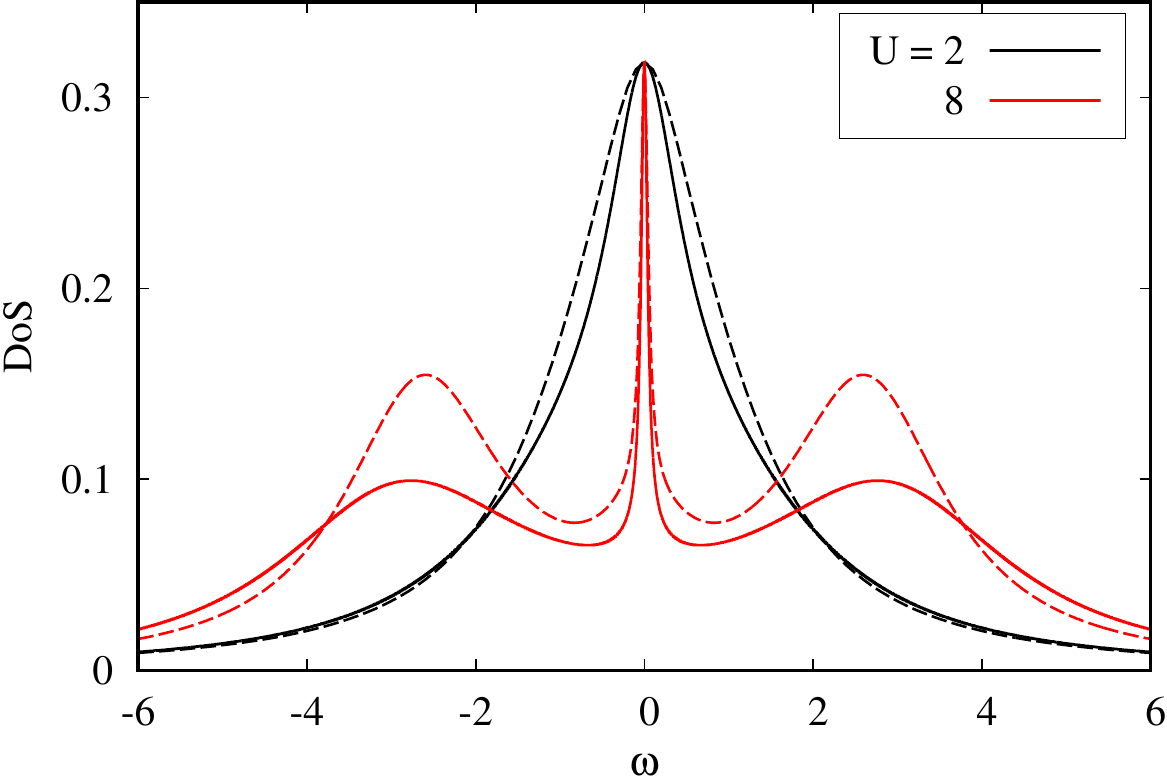}

\caption{Spectral function calculated with the full spectral self-energy, Eq.~\eqref{eq:Sigma-sp-0} (solid line), and with the asymptotic one, Eq.~\eqref{eq:SE-asympt} (dashed line), for weak and strong interaction in energy units $\Delta$. \label{fig:dos-asymp}}
\end{figure}
\begin{figure}
\includegraphics[width=8.5cm]{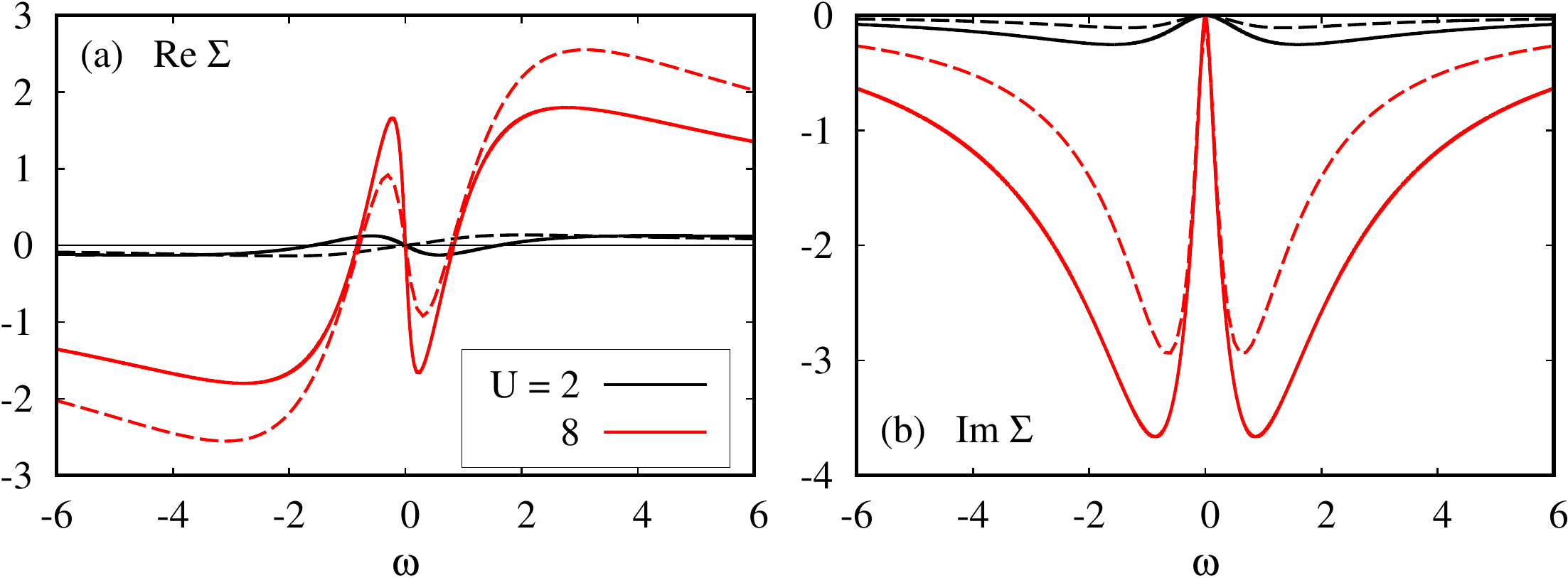}

\caption{Real (left panel) and imaginary (right panel) parts of the full (solid line) and asymptotic (dashed line) self-energies for weak and strong interaction in energy units $\Delta$. \label{fig:se-asymp}}
\end{figure}
It is convenient, in particular in more complex models, to have as simple equations in the strong-coupling regime as possible. The asymptotic representation for the self-energy, Eq.~\eqref{eq:SE-asympt}, can do the job and it is capable to deliver the qualitatively correct three-peak spectral function, see Fig.~\ref{fig:dos-asymp}. There is a good agreement in weak coupling with the full solution. The width of the central peak is asymptotically correct and only the satellite Hubbard bands are more pronounced compared to the full solution. The real and imaginary parts of the self-energy are plotted in Fig.~\ref{fig:se-asymp}. Since a one-particle self-consistency is missing, the real part of the self-energy at weak coupling does not reproduce correctly the negative slope at the Fermi energy.   

\begin{figure}
\includegraphics[width=7.5cm]{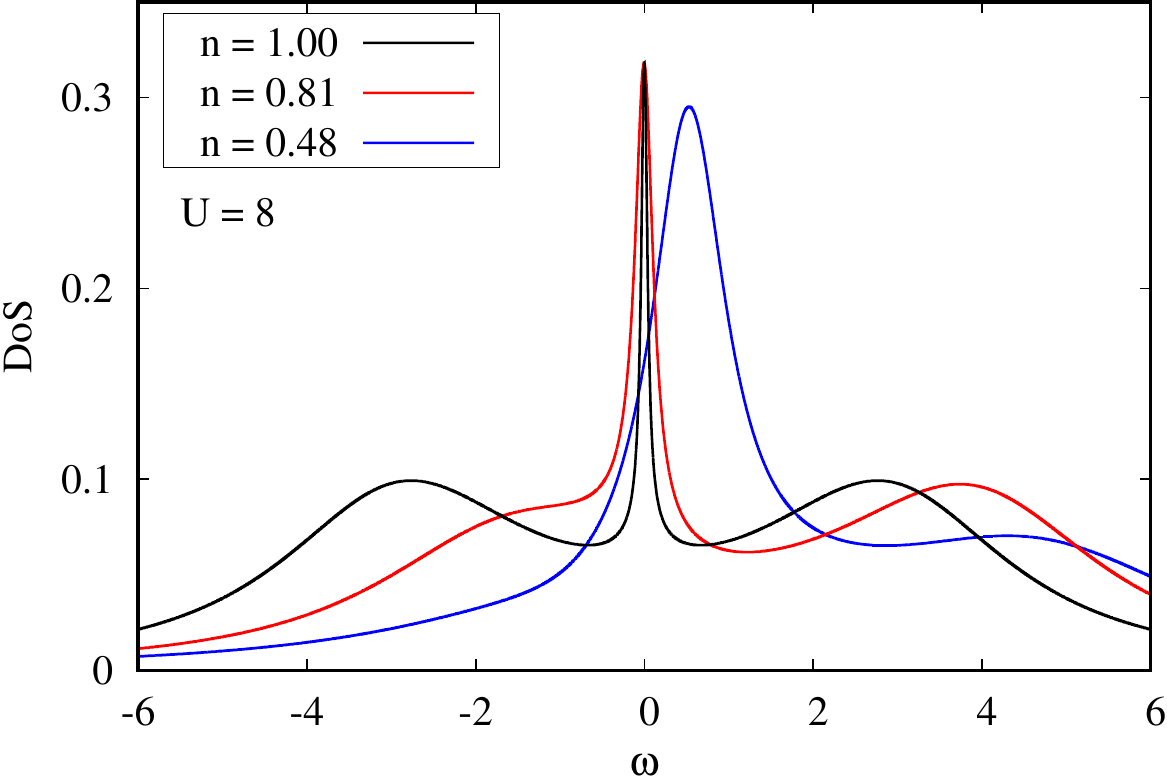}

\caption{Spectral function away from half filling in energy units $\Delta$. for values of the normalized chemical potential $\mu - \frac U2 = 0, -2\Delta$ and $-4\Delta$, respectively.   \label{fig:Sp-mu}}
\end{figure}
When we move away from half filling the central peak slowly moves away from the Fermi energy as well as the lower Hubbard band (for occupation $n<1$) moves towards the central one and eventually merges with it, Fig.~\ref{fig:Sp-mu}. It is the behavior discussed in more details in Ref.~\cite{Janis:2017ab}.  

\begin{figure}
\includegraphics[width=7.5cm]{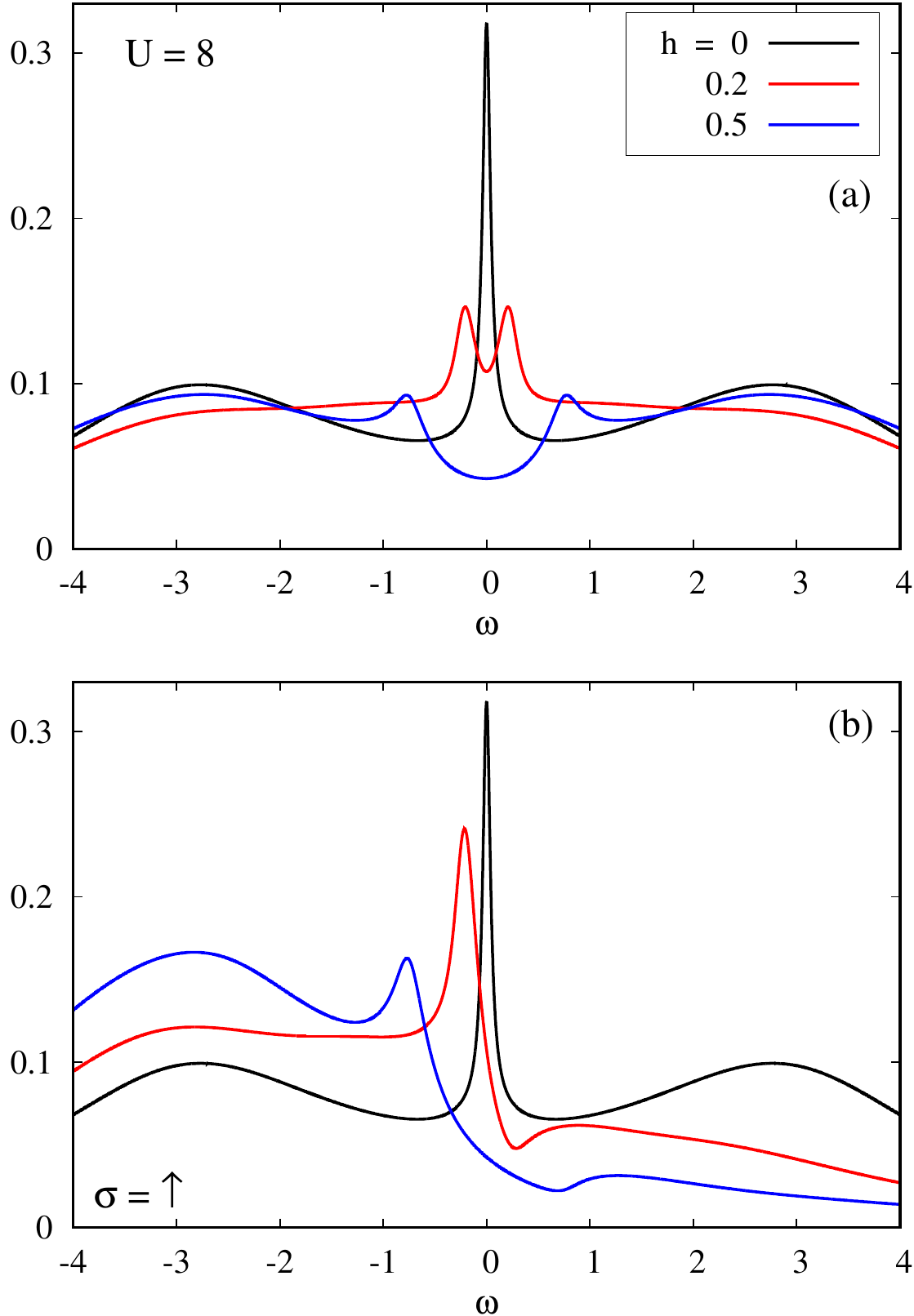}

\caption{The full spectral function (upper panel) and for spin up (lower panel) in the external magnetic field in energy units $\Delta$. \label{fig:Sp-hfull}}
\end{figure}
\begin{figure}
\includegraphics[width=8.5cm]{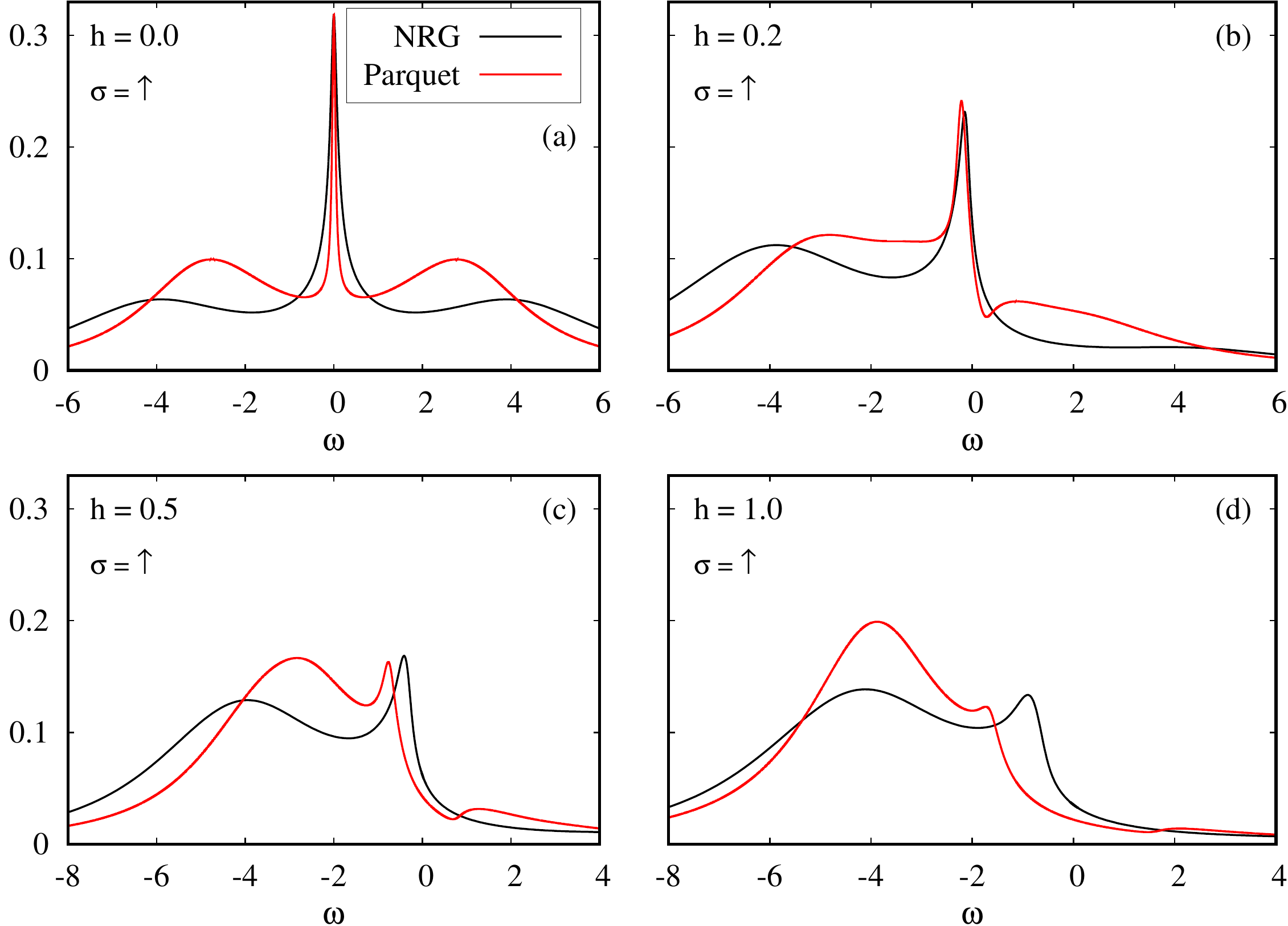}

\caption{Spectral function for spin up compared with the NRG result for several strengths of the magnetic field in energy units $\Delta$. \label{fig:Sp-hNRG}}
\end{figure}
A more interesting situation is when we break the spin-reflection symmetry, the case not discussed in our earlier publications. The magnetic field affects the Kondo strong-coupling behavior more significantly than the shift of the chemical potential from half filling. We show in Fig.~\ref{fig:Sp-hfull} the full spectral function for several values of the magnetic field. We can see that already weak magnetic fields of order $h=0.2\Delta$ split the central peak into two separate ones and lower the height of the split peaks. The movement of the central peak in the magnetic field can be better  demonstrated in the spectral function of the majority spin (up). We compared the spectral function for the majority spin with NRG in Fig.~\ref{fig:Sp-hNRG}. Qualitative features of the spectral function are well reproduced in the static approximation with the thermodynamic propagator. A dip observed at weak and intermediate field is a consequence of insufficient one-particle self-consistency.  

\begin{figure}
\includegraphics[width=8.5cm]{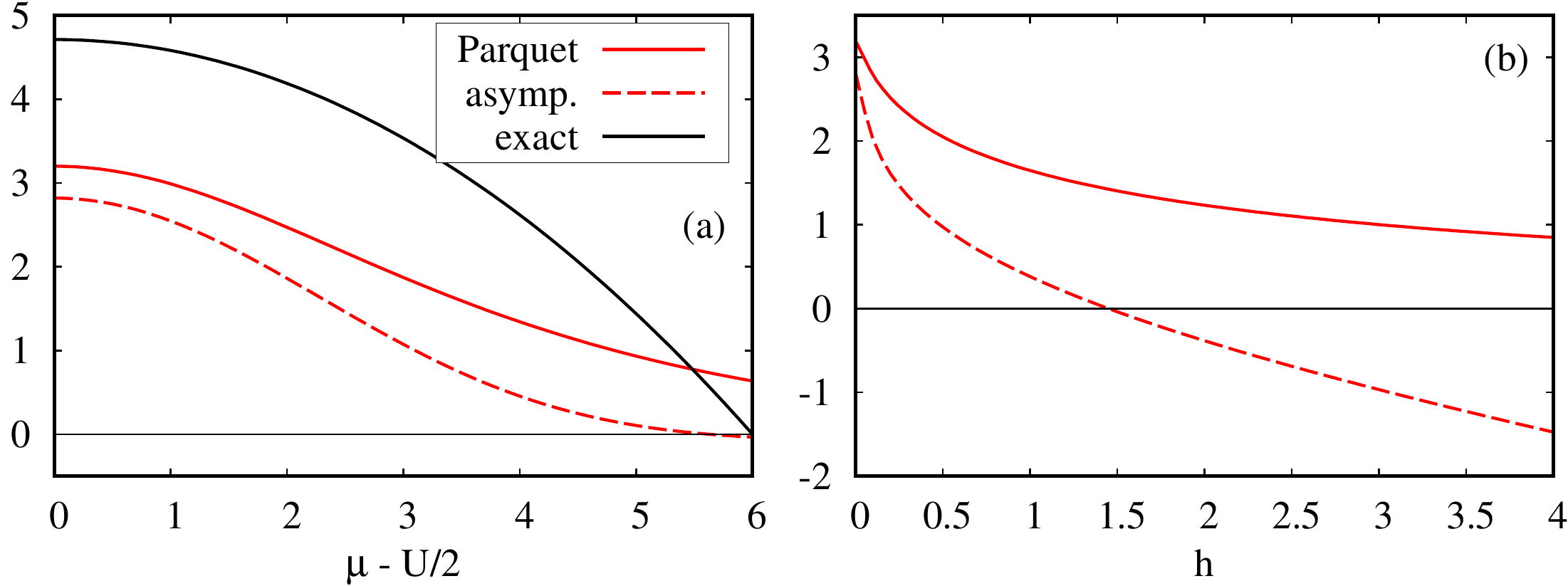}

\caption{Negative logarithm of the dimensionless Kondo scale $a_{K}$ calculated from its definition, Eq.~\eqref{eq:Det-zero} (Parquet) and from the asymptotic form, Eq.~\eqref{eq:Kondo-log}, as a function of the normalized chemical potential at zero magnetic field (left panel) and a function of the magnetic field at half filling (right panel) in energy units $\Delta$. The Bethe-ansatz result (exact) was added.  \label{fig:Kondo-mu}}
\end{figure}
The conclusion that the magnetic field moves the solution from the strong-coupling regime at half filling faster than the normalized chemical potential $\mu - U/2$ can be demonstrated in the dependence of the negative logarithm of the Kondo scale at criticality, Eq.~\eqref{eq:Kondo-log}, on both variables, Fig.~\ref{fig:Kondo-mu}. The strong-coupling regime ends where the asymptotic result goes through zero.  It happens for much smaller values of the magnetic field than for the normalized chemical potential. The difference between the initial value at $\mu -U/2=0$ of the static solution and the numerically exact one from NRG is due to the fact that $U=12 \Delta$ is not yet the true Kondo regime, since the asymptotic limit  $UD_{0} \gg  1$ in the asymptotic form in Eq.~\eqref{eq:Kondo-log} has not yet been reached. Moreover, the exact value in the limit $U\to\infty$ is $\pi/8$ while the static solution gives $1/\pi$. It is interesting to notice that the bound for the strong-coupling regime from the static solution agrees well with the exact expression for the Kondo scale. There is naturally no boundary between strong  and weak coupling in the full solution.          

\begin{figure}
\includegraphics[width=8.5cm]{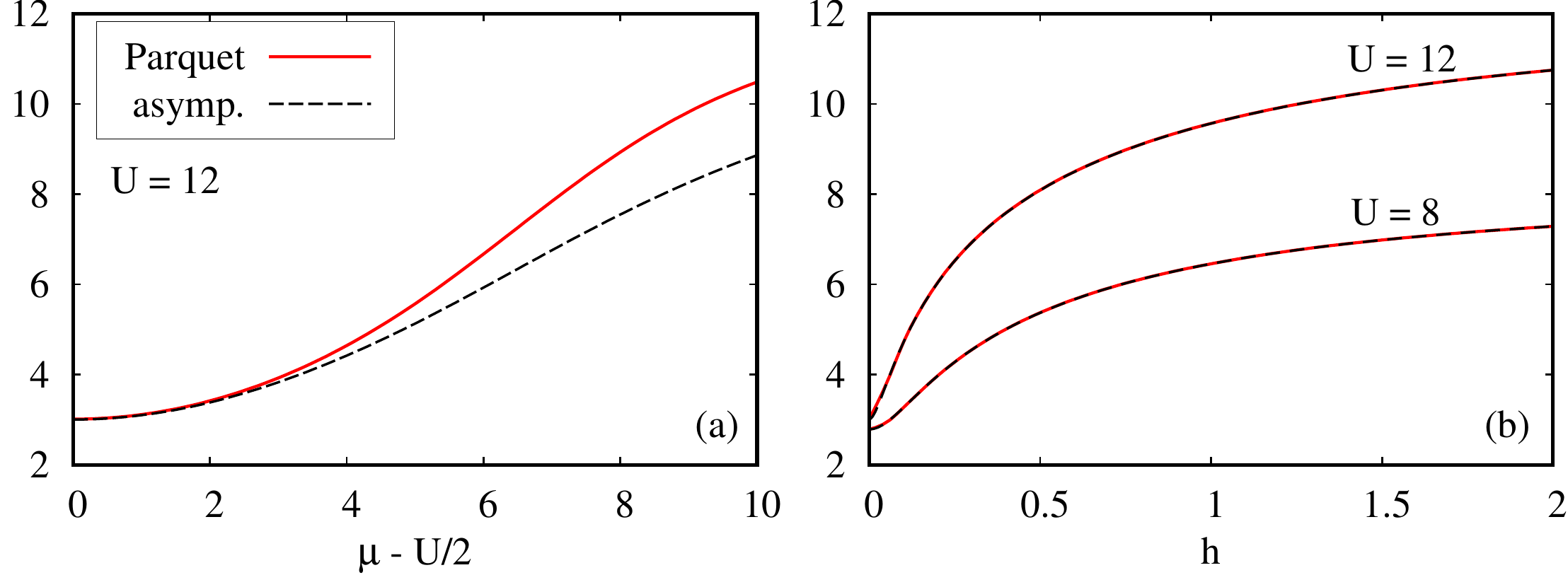}

\caption{Irreducible vertex from the electron-hole channel (effective interaction) as a function of the normalized chemical potential (left panel) and of the magnetic field (right panel)  in energy units $\Delta$. Both full (solid line) and asymptotic (dashed line) solutions from Eqs.~\eqref{eq:Lambda-zeroT} and ~\eqref{eq:Lambda-asymptotic}, respectively are plotted.  \label{fig:Lambda-muh}}
\end{figure}
The extent to which quantum fluctuations are relevant can be measured by the strength of the renormalization of the bare interaction. That is, how much the irreducible vertex $\Lambda$, effective interaction, differs from the bare one. We plotted in Fig.~\ref{fig:Lambda-muh} the dependence of $\Lambda_{\uparrow\downarrow}$ on the normalized chemical potential and magnetic field. We can again observe that the renormalization decreases faster with the increasing magnetic field than with the chemical potential. The effective interaction approaches the bare one in the weak-coupling limit with decaying quantum fluctuations.  It is surprising how well the asymptotic form reproduces the full one in the spin-polarized state.    

\section{Conclusions}

The major desired asset of the mean-field theory is its relative simplicity that allows for the analytic control of the critical behavior. It is much easier to achieve this in classical many-body models than in quantum ones. The difference is made by the dynamics brought in by quantum fluctuations in strong coupling. Although the dynamical fluctuations can be studied within a local, dynamical mean-field theory, the only accessible numerical solution does not allow for the analytic control. Further approximations are needed.  We presented in this paper a reduction scheme leading to a class of analytic mean-field theories of quantum fluctuations in strongly correlated electron systems. The resulting theories are thermodynamically consistent and conserving. They reconcile the thermodynamic Ward identity and the dynamical Schwinger-Dyson equation and are free of spurious transitions and unphysical behavior of the Hartree weak-coupling theory.  

The central objects to be determined form the diagrammatic perturbation theory of the present construction are two-particle vertex functions instead of the one-particle self-energy. The reason for that is to achieve a two-particle self-consistency that guarantees that only integrable singularities and physical phase transitions can exist. The two-particle self-consistency is derived from the parquet equations for two-particle irreducible vertices. The analytic structure of the two-particle vertices is much more complex than that of the self-energy.  A set of simplifications must be introduced to reach analytically controlled approximations.  

There are three main steps in reaching a tractable analytic mean-field theory with a two-particle self-consistency. First, one has to reduce the full set of parquet equations in order to get rid of super-divergent terms generated by mixing of different scattering channels beyond the random-phase approximation. They are assumed to be canceled by higher-order contributions beyond the two-channel parquet equations with the bare interaction as the completely irreducible vertex. If not suppressed they would prevent reaching the quantum critical behavior. Second, one moves into the critical region of the Bethe-Salpeter equation where the relevant critical fluctuations of bosonic degrees of freedom are decoupled from the non-critical ones of the fermionic degrees of freedom. It is done in analogy with the renormalization group where only small frequencies controlling criticality are explicitly considered. This leads to an explicit form of the non-critical irreducible vertex from the singular Bethe-Salpeter equation. Third, convolutions of non-critical fermionic frequencies in the Bethe-Salpeter equation are decoupled in analogy with with the mean-value theorem. This may be done either fully dynamically or statically. Here we elaborated the static approximation leading to a Hartree-like approximation with a renormalized effective interaction.  

The mean-field theory for vertex functions is not complete unless it determines the self-energy renormalizing the one-electron propagators used in the equations for the vertex functions. The consistent and conserving theory must properly relate the one and two-particle functions. We used the symmetry with respect to the external field conjugate to the order parameter of the new phase beyond the critical point of the Bethe-Salpeter equation. The normal part of the self-energy, with even symmetry with respect to the symmetry-breaking field, is determined from the Schwinger-Dyson equation using the two-particle vertex obtained from the parquet equations. The anomalous part, with odd symmetry with respect to the symmetry-breaking field, is determined from the Ward identity and the irreducible vertex from the parquet equations. The Ward identity and the Schwinger-Dyson equations connecting  the self-energy and the two-particle vertex in different ways are reconciled  in these approximations. Moreover, our approach allows for a flexible handling of the one-particle self-consistency, from non-self-consistent to fully self-consistent one-electron propagators to optimize the quantitative output.

The resulting mean-field theory with a two-particle self-consistency even in its simplest static version gives a qualitatively correct and thermodynamically consistent description of quantum criticality in SIAM. Although it is justified in the critical region of the singularities in the Bethe-Salpeter equations it can be extended to non-critical regions as well as to the ordered phase. It hence  offers a qualitative picture of the transition from weak to strong coupling as well as from low to high temperatures in models of correlated electrons. The general construction can also be used beyond the local approximation to describe the low-temperature behavior of the low-dimensional models with no long-range order at non-zero temperatures.               

\section*{Acknowledgments}
 VJ, PZ, and AK were supported by Grant No. 19-13525S of the Czech Science Foundation. VP was supported by grant INTER-COST LTC19045.
 %
%\newpage  

%\bibliographystyle{apsrev}
%\bibliography{/Users/Vaclav/Dropbox/TeX/BibTeX/parquets_MB,/Users/Vaclav/Dropbox/TeX/BibTeX/Mean_Field_Approx,/Users/Vaclav/Dropbox/TeX/BibTeX/Impurity_solver-1,/Users/Vaclav/Dropbox/TeX/BibTeX/CPA,/Users/Vaclav/Dropbox/TeX/BibTeX/Spin-glasses1}

\begin{thebibliography}{64}
\expandafter\ifx\csname natexlab\endcsname\relax\def\natexlab#1{#1}\fi
\expandafter\ifx\csname bibnamefont\endcsname\relax
  \def\bibnamefont#1{#1}\fi
\expandafter\ifx\csname bibfnamefont\endcsname\relax
  \def\bibfnamefont#1{#1}\fi
\expandafter\ifx\csname citenamefont\endcsname\relax
  \def\citenamefont#1{#1}\fi
\expandafter\ifx\csname url\endcsname\relax
  \def\url#1{\texttt{#1}}\fi
\expandafter\ifx\csname urlprefix\endcsname\relax\def\urlprefix{URL }\fi
\providecommand{\bibinfo}[2]{#2}
\providecommand{\eprint}[2][]{\url{#2}}

\bibitem[{\citenamefont{Landau}(1937)}]{Landau:1937}
\bibinfo{author}{\bibfnamefont{L.}~\bibnamefont{Landau}}, \bibinfo{journal}{Zh.
  Eksp. Teor. Fiz.} \textbf{\bibinfo{volume}{7}}, \bibinfo{pages}{19}
  (\bibinfo{year}{1937}),
  \urlprefix\url{http://www.ujp.bitp.kiev.ua/files/journals/53/si/53SI08p.pdf}.

\bibitem[{\citenamefont{Brout}(1959)}]{Brout:1959aa}
\bibinfo{author}{\bibfnamefont{R.}~\bibnamefont{Brout}},
  \bibinfo{journal}{Physical Review} \textbf{\bibinfo{volume}{115}},
  \bibinfo{pages}{824} (\bibinfo{year}{1959}).

\bibitem[{\citenamefont{Brout}(1960)}]{Brout:1960aa}
\bibinfo{author}{\bibfnamefont{R.}~\bibnamefont{Brout}},
  \bibinfo{journal}{Physical Review} \textbf{\bibinfo{volume}{118}},
  \bibinfo{pages}{1009} (\bibinfo{year}{1960}).

\bibitem[{\citenamefont{Horwitz and Callen}(1961)}]{Horwitz:1961aa}
\bibinfo{author}{\bibfnamefont{G.}~\bibnamefont{Horwitz}} \bibnamefont{and}
  \bibinfo{author}{\bibfnamefont{H.~B.} \bibnamefont{Callen}},
  \bibinfo{journal}{Physical Review} \textbf{\bibinfo{volume}{124}},
  \bibinfo{pages}{1757} (\bibinfo{year}{1961}).

\bibitem[{\citenamefont{Englert}(1963)}]{Englert:1963aa}
\bibinfo{author}{\bibfnamefont{F.}~\bibnamefont{Englert}},
  \bibinfo{journal}{Physical Review} \textbf{\bibinfo{volume}{129}},
  \bibinfo{pages}{567} (\bibinfo{year}{1963}).

\bibitem[{\citenamefont{Wilson}(1971{\natexlab{a}})}]{Wilson:1971aa}
\bibinfo{author}{\bibfnamefont{K.~G.} \bibnamefont{Wilson}},
  \bibinfo{journal}{Physical Review B} \textbf{\bibinfo{volume}{4}},
  \bibinfo{pages}{3174} (\bibinfo{year}{1971}{\natexlab{a}}).

\bibitem[{\citenamefont{Wilson}(1971{\natexlab{b}})}]{Wilson:1971ab}
\bibinfo{author}{\bibfnamefont{K.~G.} \bibnamefont{Wilson}},
  \bibinfo{journal}{Physical Review B} \textbf{\bibinfo{volume}{4}},
  \bibinfo{pages}{3184} (\bibinfo{year}{1971}{\natexlab{b}}).

\bibitem[{\citenamefont{Fisher and Gaunt}(1964)}]{Fisher:1964aa}
\bibinfo{author}{\bibfnamefont{M.~E.} \bibnamefont{Fisher}} \bibnamefont{and}
  \bibinfo{author}{\bibfnamefont{D.~S.} \bibnamefont{Gaunt}},
  \bibinfo{journal}{Physical Review} \textbf{\bibinfo{volume}{133}},
  \bibinfo{pages}{A224} (\bibinfo{year}{1964}).

\bibitem[{\citenamefont{Thompson}(1974)}]{Thompson:1974aa}
\bibinfo{author}{\bibfnamefont{C.~J.} \bibnamefont{Thompson}},
  \bibinfo{journal}{Communications in Mathematical Physics}
  \textbf{\bibinfo{volume}{36}}, \bibinfo{pages}{255} (\bibinfo{year}{1974}).

\bibitem[{\citenamefont{Sherrington and
  Kirkpatrick}(1975)}]{Sherrington:1975aa}
\bibinfo{author}{\bibfnamefont{D.}~\bibnamefont{Sherrington}} \bibnamefont{and}
  \bibinfo{author}{\bibfnamefont{S.}~\bibnamefont{Kirkpatrick}},
  \bibinfo{journal}{Physical Review Letters} \textbf{\bibinfo{volume}{35}},
  \bibinfo{pages}{1792} (\bibinfo{year}{1975}).

\bibitem[{\citenamefont{Metzner and Vollhardt}(1989)}]{Metzner:1989aa}
\bibinfo{author}{\bibfnamefont{W.}~\bibnamefont{Metzner}} \bibnamefont{and}
  \bibinfo{author}{\bibfnamefont{D.}~\bibnamefont{Vollhardt}},
  \bibinfo{journal}{Physical Review Letters} \textbf{\bibinfo{volume}{62}},
  \bibinfo{pages}{324} (\bibinfo{year}{1989}).

\bibitem[{\citenamefont{Jani{\v s}}(1989)}]{Janis:1989aa}
\bibinfo{author}{\bibfnamefont{V.}~\bibnamefont{Jani{\v s}}},
  \bibinfo{journal}{Physical Review B} \textbf{\bibinfo{volume}{40}},
  \bibinfo{pages}{11331} (\bibinfo{year}{1989}).

\bibitem[{\citenamefont{Brandt and Mielsch}(1989)}]{Brandt:1989aa}
\bibinfo{author}{\bibfnamefont{U.}~\bibnamefont{Brandt}} \bibnamefont{and}
  \bibinfo{author}{\bibfnamefont{C.}~\bibnamefont{Mielsch}},
  \bibinfo{journal}{Zeitschrift f{\"u}r Physik B Condensed Matter}
  \textbf{\bibinfo{volume}{75}}, \bibinfo{pages}{365} (\bibinfo{year}{1989}).

\bibitem[{\citenamefont{Jani{\v s}}(1991)}]{Janis:1991aa}
\bibinfo{author}{\bibfnamefont{V.}~\bibnamefont{Jani{\v s}}},
  \bibinfo{journal}{Zeitschrift f{\"u}r Physik B Condensed Matter}
  \textbf{\bibinfo{volume}{83}}, \bibinfo{pages}{227} (\bibinfo{year}{1991}).

\bibitem[{\citenamefont{Georges et~al.}(1996)\citenamefont{Georges, Kotliar,
  Krauth, and Rozenberg}}]{Georges:1996aa}
\bibinfo{author}{\bibfnamefont{A.}~\bibnamefont{Georges}},
  \bibinfo{author}{\bibfnamefont{G.}~\bibnamefont{Kotliar}},
  \bibinfo{author}{\bibfnamefont{W.}~\bibnamefont{Krauth}}, \bibnamefont{and}
  \bibinfo{author}{\bibfnamefont{M.~J.} \bibnamefont{Rozenberg}},
  \bibinfo{journal}{Reviews of Modern Physics} \textbf{\bibinfo{volume}{68}},
  \bibinfo{pages}{13} (\bibinfo{year}{1996}).

\bibitem[{\citenamefont{Kotliar et~al.}(2006)\citenamefont{Kotliar, Savrasov,
  Haule, Oudovenko, Parcollet, and Marionetti}}]{Kotliar:2006aa}
\bibinfo{author}{\bibfnamefont{G.}~\bibnamefont{Kotliar}},
  \bibinfo{author}{\bibfnamefont{S.~Y.} \bibnamefont{Savrasov}},
  \bibinfo{author}{\bibfnamefont{K.}~\bibnamefont{Haule}},
  \bibinfo{author}{\bibfnamefont{V.~D.} \bibnamefont{Oudovenko}},
  \bibinfo{author}{\bibfnamefont{O.}~\bibnamefont{Parcollet}},
  \bibnamefont{and} \bibinfo{author}{\bibfnamefont{C.~A.}
  \bibnamefont{Marionetti}}, \bibinfo{journal}{Reviews of Modern Physics}
  \textbf{\bibinfo{volume}{78}}, \bibinfo{pages}{865} (\bibinfo{year}{2006}).

\bibitem[{\citenamefont{Rohringer et~al.}(2018)\citenamefont{Rohringer,
  Hafermann, Toschi, Katanin, Antipov, Katsnelson, Lichtenstein, Rubtsov, and
  Held}}]{Rohringer:2018aa}
\bibinfo{author}{\bibfnamefont{G.}~\bibnamefont{Rohringer}},
  \bibinfo{author}{\bibfnamefont{H.}~\bibnamefont{Hafermann}},
  \bibinfo{author}{\bibfnamefont{A.}~\bibnamefont{Toschi}},
  \bibinfo{author}{\bibfnamefont{A.}~\bibnamefont{Katanin}},
  \bibinfo{author}{\bibfnamefont{A.}~\bibnamefont{Antipov}},
  \bibinfo{author}{\bibfnamefont{M.}~\bibnamefont{Katsnelson}},
  \bibinfo{author}{\bibfnamefont{A.}~\bibnamefont{Lichtenstein}},
  \bibinfo{author}{\bibfnamefont{A.}~\bibnamefont{Rubtsov}}, \bibnamefont{and}
  \bibinfo{author}{\bibfnamefont{K.}~\bibnamefont{Held}},
  \bibinfo{journal}{Reviews of Modern Physics} \textbf{\bibinfo{volume}{90}},
  \bibinfo{pages}{025003} (\bibinfo{year}{2018}).

\bibitem[{\citenamefont{Kopietz et~al.}(2010)\citenamefont{Kopietz, Bartosch,
  and Sch{\"u}tz}}]{Kopietz:2010ab}
\bibinfo{author}{\bibfnamefont{P.}~\bibnamefont{Kopietz}},
  \bibinfo{author}{\bibfnamefont{L.}~\bibnamefont{Bartosch}}, \bibnamefont{and}
  \bibinfo{author}{\bibfnamefont{F.}~\bibnamefont{Sch{\"u}tz}},
  \emph{\bibinfo{title}{Introduction to the Functional Renormalization Group}},
  vol. \bibinfo{volume}{798} of \emph{\bibinfo{series}{Lecture Notes in
  Physics}} (\bibinfo{publisher}{Springer}, \bibinfo{address}{Berlin
  Heidelberg}, \bibinfo{year}{2010}).

\bibitem[{\citenamefont{Jani{\v s} and Augustinsk{\'y}}(2007)}]{Janis:2007aa}
\bibinfo{author}{\bibfnamefont{V.}~\bibnamefont{Jani{\v s}}} \bibnamefont{and}
  \bibinfo{author}{\bibfnamefont{P.}~\bibnamefont{Augustinsk{\'y}}},
  \bibinfo{journal}{Physical Review B} \textbf{\bibinfo{volume}{75}},
  \bibinfo{pages}{165108} (\bibinfo{year}{2007}).

\bibitem[{\citenamefont{Jani{\v s} and Augustinsk{\'y}}(2008)}]{Janis:2008ab}
\bibinfo{author}{\bibfnamefont{V.}~\bibnamefont{Jani{\v s}}} \bibnamefont{and}
  \bibinfo{author}{\bibfnamefont{P.}~\bibnamefont{Augustinsk{\'y}}},
  \bibinfo{journal}{Physical Review B} \textbf{\bibinfo{volume}{77}},
  \bibinfo{pages}{085106} (\bibinfo{year}{2008}).

\bibitem[{\citenamefont{Jani{\v s}
  et~al.}(2017{\natexlab{a}})\citenamefont{Jani{\v s}, Kauch, and
  Pokorn{\'y}}}]{Janis:2017aa}
\bibinfo{author}{\bibfnamefont{V.}~\bibnamefont{Jani{\v s}}},
  \bibinfo{author}{\bibfnamefont{A.}~\bibnamefont{Kauch}}, \bibnamefont{and}
  \bibinfo{author}{\bibfnamefont{V.}~\bibnamefont{Pokorn{\'y}}},
  \bibinfo{journal}{Physical Review B} \textbf{\bibinfo{volume}{95}},
  \bibinfo{pages}{045108} (\bibinfo{year}{2017}{\natexlab{a}}).

\bibitem[{\citenamefont{Jani{\v s}
  et~al.}(2017{\natexlab{b}})\citenamefont{Jani{\v s}, Pokorn{\'y}, and
  Kauch}}]{Janis:2017ab}
\bibinfo{author}{\bibfnamefont{V.}~\bibnamefont{Jani{\v s}}},
  \bibinfo{author}{\bibfnamefont{V.}~\bibnamefont{Pokorn{\'y}}},
  \bibnamefont{and} \bibinfo{author}{\bibfnamefont{A.}~\bibnamefont{Kauch}},
  \bibinfo{journal}{Physical Review B} \textbf{\bibinfo{volume}{95}},
  \bibinfo{pages}{165113} (\bibinfo{year}{2017}{\natexlab{b}}).

\bibitem[{\citenamefont{Hedin}(1965)}]{Hedin:1965aa}
\bibinfo{author}{\bibfnamefont{L.}~\bibnamefont{Hedin}},
  \bibinfo{journal}{Physical Review} \textbf{\bibinfo{volume}{139}},
  \bibinfo{pages}{A796} (\bibinfo{year}{1965}).

\bibitem[{\citenamefont{Aryasetiawan and
  Gunnarsson}(1997)}]{Aryasetiawan:1997aa}
\bibinfo{author}{\bibfnamefont{F.}~\bibnamefont{Aryasetiawan}}
  \bibnamefont{and}
  \bibinfo{author}{\bibfnamefont{O.}~\bibnamefont{Gunnarsson}},
  \bibinfo{journal}{Reports on Progress in Physics}
  \textbf{\bibinfo{volume}{61}}, \bibinfo{pages}{237} (\bibinfo{year}{1997}).

\bibitem[{\citenamefont{Jani{\v s}}(1998)}]{Janis:1998aa}
\bibinfo{author}{\bibfnamefont{V.}~\bibnamefont{Jani{\v s}}},
  \bibinfo{journal}{Journal of Physics: Condensed Matter}
  \textbf{\bibinfo{volume}{10}}, \bibinfo{pages}{2915} (\bibinfo{year}{1998}).

\bibitem[{\citenamefont{Jani{\v s}}(1999)}]{Janis:1999aa}
\bibinfo{author}{\bibfnamefont{V.}~\bibnamefont{Jani{\v s}}},
  \bibinfo{journal}{Physical Review B} \textbf{\bibinfo{volume}{60}},
  \bibinfo{pages}{11345} (\bibinfo{year}{1999}).

\bibitem[{\citenamefont{Toschi et~al.}(2007)\citenamefont{Toschi, Katanin, and
  Held}}]{Toschi:2007aa}
\bibinfo{author}{\bibfnamefont{A.}~\bibnamefont{Toschi}},
  \bibinfo{author}{\bibfnamefont{A.~A.} \bibnamefont{Katanin}},
  \bibnamefont{and} \bibinfo{author}{\bibfnamefont{K.}~\bibnamefont{Held}},
  \bibinfo{journal}{Physical Review B} \textbf{\bibinfo{volume}{75}},
  \bibinfo{pages}{045118} (\bibinfo{year}{2007}).

\bibitem[{\citenamefont{Rubtsov et~al.}(2008)\citenamefont{Rubtsov, Katsnelson,
  and Lichtenstein}}]{Rubtsov:2008aa}
\bibinfo{author}{\bibfnamefont{A.~N.} \bibnamefont{Rubtsov}},
  \bibinfo{author}{\bibfnamefont{M.~I.} \bibnamefont{Katsnelson}},
  \bibnamefont{and} \bibinfo{author}{\bibfnamefont{A.~I.}
  \bibnamefont{Lichtenstein}}, \bibinfo{journal}{Physical Review B}
  \textbf{\bibinfo{volume}{77}}, \bibinfo{pages}{033101}
  (\bibinfo{year}{2008}).

\bibitem[{\citenamefont{Rohringer et~al.}(2011)\citenamefont{Rohringer, Toschi,
  Katanin, and Held}}]{Rohringer:2011aa}
\bibinfo{author}{\bibfnamefont{G.}~\bibnamefont{Rohringer}},
  \bibinfo{author}{\bibfnamefont{A.}~\bibnamefont{Toschi}},
  \bibinfo{author}{\bibfnamefont{A.}~\bibnamefont{Katanin}}, \bibnamefont{and}
  \bibinfo{author}{\bibfnamefont{K.}~\bibnamefont{Held}},
  \bibinfo{journal}{Physical Review Letters} \textbf{\bibinfo{volume}{107}},
  \bibinfo{pages}{256402} (\bibinfo{year}{2011}).

\bibitem[{\citenamefont{Rubtsov et~al.}(2012)\citenamefont{Rubtsov, Katsnelson,
  and Lichtenstein}}]{Rubtsov:2012aa}
\bibinfo{author}{\bibfnamefont{A.}~\bibnamefont{Rubtsov}},
  \bibinfo{author}{\bibfnamefont{M.}~\bibnamefont{Katsnelson}},
  \bibnamefont{and}
  \bibinfo{author}{\bibfnamefont{A.}~\bibnamefont{Lichtenstein}},
  \bibinfo{journal}{Annals of Physics} \textbf{\bibinfo{volume}{327}},
  \bibinfo{pages}{1320 } (\bibinfo{year}{2012}).

\bibitem[{\citenamefont{Rohringer et~al.}(2013)\citenamefont{Rohringer, Toschi,
  Hafermann, Held, Anisimov, and Katanin}}]{Rohringer:2013aa}
\bibinfo{author}{\bibfnamefont{G.}~\bibnamefont{Rohringer}},
  \bibinfo{author}{\bibfnamefont{A.}~\bibnamefont{Toschi}},
  \bibinfo{author}{\bibfnamefont{H.}~\bibnamefont{Hafermann}},
  \bibinfo{author}{\bibfnamefont{K.}~\bibnamefont{Held}},
  \bibinfo{author}{\bibfnamefont{V.~I.} \bibnamefont{Anisimov}},
  \bibnamefont{and} \bibinfo{author}{\bibfnamefont{A.~A.}
  \bibnamefont{Katanin}}, \bibinfo{journal}{Physical Review B}
  \textbf{\bibinfo{volume}{88}}, \bibinfo{pages}{115112}
  (\bibinfo{year}{2013}).

\bibitem[{\citenamefont{Valli et~al.}(2015)\citenamefont{Valli, Sch{\"a}fer,
  Thunstr{\"o}m, Rohringer, Andergassen, Sangiovanni, Held, and
  Toschi}}]{Valli:2015aa}
\bibinfo{author}{\bibfnamefont{A.}~\bibnamefont{Valli}},
  \bibinfo{author}{\bibfnamefont{T.}~\bibnamefont{Sch{\"a}fer}},
  \bibinfo{author}{\bibfnamefont{P.}~\bibnamefont{Thunstr{\"o}m}},
  \bibinfo{author}{\bibfnamefont{G.}~\bibnamefont{Rohringer}},
  \bibinfo{author}{\bibfnamefont{S.}~\bibnamefont{Andergassen}},
  \bibinfo{author}{\bibfnamefont{G.}~\bibnamefont{Sangiovanni}},
  \bibinfo{author}{\bibfnamefont{K.}~\bibnamefont{Held}}, \bibnamefont{and}
  \bibinfo{author}{\bibfnamefont{A.}~\bibnamefont{Toschi}},
  \bibinfo{journal}{Physical Review B} \textbf{\bibinfo{volume}{91}},
  \bibinfo{pages}{115115} (\bibinfo{year}{2015}).

\bibitem[{\citenamefont{Hirschmeier et~al.}(2015)\citenamefont{Hirschmeier,
  Hafermann, Gull, Lichtenstein, and Antipov}}]{Hirschmeier:2015aa}
\bibinfo{author}{\bibfnamefont{D.}~\bibnamefont{Hirschmeier}},
  \bibinfo{author}{\bibfnamefont{H.}~\bibnamefont{Hafermann}},
  \bibinfo{author}{\bibfnamefont{E.}~\bibnamefont{Gull}},
  \bibinfo{author}{\bibfnamefont{A.~I.} \bibnamefont{Lichtenstein}},
  \bibnamefont{and} \bibinfo{author}{\bibfnamefont{A.~E.}
  \bibnamefont{Antipov}}, \bibinfo{journal}{Physical Review B}
  \textbf{\bibinfo{volume}{92}}, \bibinfo{pages}{144409}
  (\bibinfo{year}{2015}).

\bibitem[{\citenamefont{Ayral and Parcollet}(2016)}]{Ayral:2016aa}
\bibinfo{author}{\bibfnamefont{T.}~\bibnamefont{Ayral}} \bibnamefont{and}
  \bibinfo{author}{\bibfnamefont{O.}~\bibnamefont{Parcollet}},
  \bibinfo{journal}{Physical Review B} \textbf{\bibinfo{volume}{94}},
  \bibinfo{pages}{075159} (\bibinfo{year}{2016}).

\bibitem[{\citenamefont{Kugler and Delft}(2018)}]{Kugler:2018aa}
\bibinfo{author}{\bibfnamefont{F.~B.} \bibnamefont{Kugler}} \bibnamefont{and}
  \bibinfo{author}{\bibfnamefont{J.~v.} \bibnamefont{Delft}},
  \bibinfo{journal}{New Journal of Physics} \textbf{\bibinfo{volume}{20}},
  \bibinfo{pages}{123029} (\bibinfo{year}{2018}).

\bibitem[{\citenamefont{Del~Re et~al.}(2019)\citenamefont{Del~Re, Capone, and
  Toschi}}]{Del_Re:2019aa}
\bibinfo{author}{\bibfnamefont{L.}~\bibnamefont{Del~Re}},
  \bibinfo{author}{\bibfnamefont{M.}~\bibnamefont{Capone}}, \bibnamefont{and}
  \bibinfo{author}{\bibfnamefont{A.}~\bibnamefont{Toschi}},
  \bibinfo{journal}{Physical Review B} \textbf{\bibinfo{volume}{99}},
  \bibinfo{pages}{045137} (\bibinfo{year}{2019}).

\bibitem[{\citenamefont{Yang et~al.}(2009)\citenamefont{Yang, Fotso, Liu,
  Maier, Tomko, D'Azevedo, Scalettar, Pruschke, and Jarrell}}]{Yang:2009aa}
\bibinfo{author}{\bibfnamefont{S.~X.} \bibnamefont{Yang}},
  \bibinfo{author}{\bibfnamefont{H.}~\bibnamefont{Fotso}},
  \bibinfo{author}{\bibfnamefont{J.}~\bibnamefont{Liu}},
  \bibinfo{author}{\bibfnamefont{T.~A.} \bibnamefont{Maier}},
  \bibinfo{author}{\bibfnamefont{K.}~\bibnamefont{Tomko}},
  \bibinfo{author}{\bibfnamefont{E.~F.} \bibnamefont{D'Azevedo}},
  \bibinfo{author}{\bibfnamefont{R.~T.} \bibnamefont{Scalettar}},
  \bibinfo{author}{\bibfnamefont{T.}~\bibnamefont{Pruschke}}, \bibnamefont{and}
  \bibinfo{author}{\bibfnamefont{M.}~\bibnamefont{Jarrell}},
  \bibinfo{journal}{Physical Review E} \textbf{\bibinfo{volume}{80}},
  \bibinfo{pages}{046706} (\bibinfo{year}{2009}).

\bibitem[{\citenamefont{Tam et~al.}(2013)\citenamefont{Tam, Fotso, Yang, Lee,
  Moreno, Ramanujam, and Jarrell}}]{Tam:2013aa}
\bibinfo{author}{\bibfnamefont{K.-M.} \bibnamefont{Tam}},
  \bibinfo{author}{\bibfnamefont{H.}~\bibnamefont{Fotso}},
  \bibinfo{author}{\bibfnamefont{S.-X.} \bibnamefont{Yang}},
  \bibinfo{author}{\bibfnamefont{T.-W.} \bibnamefont{Lee}},
  \bibinfo{author}{\bibfnamefont{J.}~\bibnamefont{Moreno}},
  \bibinfo{author}{\bibfnamefont{J.}~\bibnamefont{Ramanujam}},
  \bibnamefont{and} \bibinfo{author}{\bibfnamefont{M.}~\bibnamefont{Jarrell}},
  \bibinfo{journal}{Physical Review E} \textbf{\bibinfo{volume}{87}},
  \bibinfo{pages}{013311} (\bibinfo{year}{2013}).

\bibitem[{\citenamefont{Li et~al.}(2016)\citenamefont{Li, Wentzell, Pudleiner,
  Thunstr{\"o}m, and Held}}]{Li:2016aa}
\bibinfo{author}{\bibfnamefont{G.}~\bibnamefont{Li}},
  \bibinfo{author}{\bibfnamefont{N.}~\bibnamefont{Wentzell}},
  \bibinfo{author}{\bibfnamefont{P.}~\bibnamefont{Pudleiner}},
  \bibinfo{author}{\bibfnamefont{P.}~\bibnamefont{Thunstr{\"o}m}},
  \bibnamefont{and} \bibinfo{author}{\bibfnamefont{K.}~\bibnamefont{Held}},
  \bibinfo{journal}{Physical Review B} \textbf{\bibinfo{volume}{93}},
  \bibinfo{pages}{165103} (\bibinfo{year}{2016}).

\bibitem[{\citenamefont{Vilk and Tremblay}(1997)}]{Vilk:1997aa}
\bibinfo{author}{\bibfnamefont{Y.}~\bibnamefont{Vilk}} \bibnamefont{and}
  \bibinfo{author}{\bibfnamefont{A.-M.~S.} \bibnamefont{Tremblay}},
  \bibinfo{journal}{Journal de Physique} \textbf{\bibinfo{volume}{7}},
  \bibinfo{pages}{1309} (\bibinfo{year}{1997}).

\bibitem[{\citenamefont{Kusunose}(2010)}]{Kusunose:2010aa}
\bibinfo{author}{\bibfnamefont{H.}~\bibnamefont{Kusunose}},
  \bibinfo{journal}{Journal of the Physical Society of Japan}
  \textbf{\bibinfo{volume}{79}}, \bibinfo{pages}{094707}
  (\bibinfo{year}{2010}).

\bibitem[{\citenamefont{Baym and Kadanoff}(1961)}]{Baym:1961aa}
\bibinfo{author}{\bibfnamefont{G.}~\bibnamefont{Baym}} \bibnamefont{and}
  \bibinfo{author}{\bibfnamefont{L.~P.} \bibnamefont{Kadanoff}},
  \bibinfo{journal}{Physical Review} \textbf{\bibinfo{volume}{124}},
  \bibinfo{pages}{287} (\bibinfo{year}{1961}).

\bibitem[{\citenamefont{Baym}(1962)}]{Baym:1962aa}
\bibinfo{author}{\bibfnamefont{G.}~\bibnamefont{Baym}},
  \bibinfo{journal}{Physical Review} \textbf{\bibinfo{volume}{127}},
  \bibinfo{pages}{1391} (\bibinfo{year}{1962}).

\bibitem[{\citenamefont{Dominicis and
  Martin}(1964{\natexlab{a}})}]{DeDominicis:1964aa}
\bibinfo{author}{\bibfnamefont{C.~D.} \bibnamefont{Dominicis}}
  \bibnamefont{and} \bibinfo{author}{\bibfnamefont{P.~C.}
  \bibnamefont{Martin}}, \bibinfo{journal}{Journal of Mathematical Physics}
  \textbf{\bibinfo{volume}{5}}, \bibinfo{pages}{14}
  (\bibinfo{year}{1964}{\natexlab{a}}).

\bibitem[{\citenamefont{Dominicis and
  Martin}(1964{\natexlab{b}})}]{DeDominicis:1964ab}
\bibinfo{author}{\bibfnamefont{C.~D.} \bibnamefont{Dominicis}}
  \bibnamefont{and} \bibinfo{author}{\bibfnamefont{P.~C.}
  \bibnamefont{Martin}}, \bibinfo{journal}{Journal of Mathematical Physics}
  \textbf{\bibinfo{volume}{5}}, \bibinfo{pages}{31}
  (\bibinfo{year}{1964}{\natexlab{b}}).

\bibitem[{\citenamefont{Roulet et~al.}(1969)\citenamefont{Roulet, Gavoret, and
  Nozieres}}]{Roulet:1969aa}
\bibinfo{author}{\bibfnamefont{B.}~\bibnamefont{Roulet}},
  \bibinfo{author}{\bibfnamefont{J.}~\bibnamefont{Gavoret}}, \bibnamefont{and}
  \bibinfo{author}{\bibfnamefont{P.}~\bibnamefont{Nozieres}},
  \bibinfo{journal}{Physical Review} \textbf{\bibinfo{volume}{178}},
  \bibinfo{pages}{1072} (\bibinfo{year}{1969}).

\bibitem[{\citenamefont{Weiner}(1970)}]{Weiner:1970aa}
\bibinfo{author}{\bibfnamefont{R.~A.} \bibnamefont{Weiner}},
  \bibinfo{journal}{Physical Review Letters} \textbf{\bibinfo{volume}{24}},
  \bibinfo{pages}{1071} (\bibinfo{year}{1970}).

\bibitem[{\citenamefont{Weiner}(1971)}]{Weiner:1971aa}
\bibinfo{author}{\bibfnamefont{R.~A.} \bibnamefont{Weiner}},
  \bibinfo{journal}{Physical Review B} \textbf{\bibinfo{volume}{4}},
  \bibinfo{pages}{3165} (\bibinfo{year}{1971}).

\bibitem[{\citenamefont{Bickers and White}(1991)}]{Bickers:1991aa}
\bibinfo{author}{\bibfnamefont{N.~E.} \bibnamefont{Bickers}} \bibnamefont{and}
  \bibinfo{author}{\bibfnamefont{S.~R.} \bibnamefont{White}},
  \bibinfo{journal}{Physical Review B} \textbf{\bibinfo{volume}{43}},
  \bibinfo{pages}{8044} (\bibinfo{year}{1991}).

\bibitem[{\citenamefont{Bickers and Scalapino}(1992)}]{Bickers:1992aa}
\bibinfo{author}{\bibfnamefont{N.~E.} \bibnamefont{Bickers}} \bibnamefont{and}
  \bibinfo{author}{\bibfnamefont{D.~J.} \bibnamefont{Scalapino}},
  \bibinfo{journal}{Physical Review B} \textbf{\bibinfo{volume}{46}},
  \bibinfo{pages}{8050} (\bibinfo{year}{1992}).

\bibitem[{\citenamefont{Bickers and Scalapino}(1989)}]{Bickers:1989ab}
\bibinfo{author}{\bibfnamefont{N.~E.} \bibnamefont{Bickers}} \bibnamefont{and}
  \bibinfo{author}{\bibfnamefont{D.~J.} \bibnamefont{Scalapino}},
  \bibinfo{journal}{Annals of Physics} \textbf{\bibinfo{volume}{193}},
  \bibinfo{pages}{206} (\bibinfo{year}{1989}).

\bibitem[{\citenamefont{Hamann}(1969)}]{Hamann:1969aa}
\bibinfo{author}{\bibfnamefont{D.~R.} \bibnamefont{Hamann}},
  \bibinfo{journal}{Physical Review} \textbf{\bibinfo{volume}{186}},
  \bibinfo{pages}{549} (\bibinfo{year}{1969}).

\bibitem[{\citenamefont{Chen and Bickers}(1992)}]{Chen:1992aa}
\bibinfo{author}{\bibfnamefont{C.~X.} \bibnamefont{Chen}} \bibnamefont{and}
  \bibinfo{author}{\bibfnamefont{N.~E.} \bibnamefont{Bickers}},
  \bibinfo{journal}{Solid State Communications} \textbf{\bibinfo{volume}{82}},
  \bibinfo{pages}{311} (\bibinfo{year}{1992}).

\bibitem[{\citenamefont{Jani{\v s}}(2006)}]{Janis:2006ab}
\bibinfo{author}{\bibfnamefont{V.}~\bibnamefont{Jani{\v s}}},
  \bibinfo{journal}{Condens. Matter Physics} \textbf{\bibinfo{volume}{9}},
  \bibinfo{pages}{499} (\bibinfo{year}{2006}).

\bibitem[{\citenamefont{Bickers}(1991)}]{Bickers:1991ab}
\bibinfo{author}{\bibfnamefont{N.}~\bibnamefont{Bickers}},
  \bibinfo{journal}{International Journal of Modern Physics B}
  \textbf{\bibinfo{volume}{05}}, \bibinfo{pages}{253} (\bibinfo{year}{1991}).

\bibitem[{\citenamefont{Jani{\v s}}(2009)}]{Janis:2009aa}
\bibinfo{author}{\bibfnamefont{V.}~\bibnamefont{Jani{\v s}}},
  \bibinfo{journal}{Journal of Physics: Condensed Matter}
  \textbf{\bibinfo{volume}{21}}, \bibinfo{pages}{485501}
  (\bibinfo{year}{2009}).

\bibitem[{Note1()}]{Note1}
Note1, \bibinfo{note}{the high-frequency structure of the irreducible vertex
  may be rather complex with non-analyticities \cite
  {Janis:2014aa,Schafer:2013aa,Chalupa:2018aa}. They are related to increasing
  imaginary part of the self-energy and are precursors of the metal-insulator
  transition. They may affect the low-frequency behavior only beyond the
  Fermi-liquid regime, since the imaginary part of the self-energy vanishes at
  the Fermi energy for Fermi liquids.}

\bibitem[{\citenamefont{Jani{\v s} and Kl{\'\i}{\v c}}(2019)}]{Janis:2019ab}
\bibinfo{author}{\bibfnamefont{V.}~\bibnamefont{Jani{\v s}}} \bibnamefont{and}
  \bibinfo{author}{\bibfnamefont{A.}~\bibnamefont{Kl{\'\i}{\v c}}}
  (\bibinfo{year}{2019}), \bibinfo{note}{arXiv:1909.02292 [cond-mat.str-el]}.

\bibitem[{\citenamefont{\v{Z}itko}(2014)}]{NRGLjubljana}
\bibinfo{author}{\bibfnamefont{R.}~\bibnamefont{\v{Z}itko}}
  (\bibinfo{year}{2014}), \bibinfo{note}{{h}ttp://nrgljubljana.ijs.si}.

\bibitem[{\citenamefont{Hofstetter}(2000)}]{Hofstetter:2000aa}
\bibinfo{author}{\bibfnamefont{W.}~\bibnamefont{Hofstetter}},
  \bibinfo{journal}{Physical Review Letters} \textbf{\bibinfo{volume}{85}},
  \bibinfo{pages}{1508} (\bibinfo{year}{2000}).

\bibitem[{\citenamefont{Hewson}(1993)}]{Hewson:1993aa}
\bibinfo{author}{\bibfnamefont{A.~C.} \bibnamefont{Hewson}},
  \emph{\bibinfo{title}{The Kondo Problem to Heavy Fermions}},
  vol.~\bibinfo{volume}{2} of \emph{\bibinfo{series}{Cambridge Studies in
  Magnetism}} (\bibinfo{publisher}{Cambridge University Press},
  \bibinfo{address}{Cambridge, United Kingdom}, \bibinfo{year}{1993}).

\bibitem[{\citenamefont{Jani{\v s} and Pokorn{\'y}}(2014)}]{Janis:2014aa}
\bibinfo{author}{\bibfnamefont{V.}~\bibnamefont{Jani{\v s}}} \bibnamefont{and}
  \bibinfo{author}{\bibfnamefont{V.}~\bibnamefont{Pokorn{\'y}}},
  \bibinfo{journal}{Physical Review B} \textbf{\bibinfo{volume}{90}},
  \bibinfo{pages}{045143} (\bibinfo{year}{2014}).

\bibitem[{\citenamefont{Sch{\"a}fer et~al.}(2013)\citenamefont{Sch{\"a}fer,
  Rohringer, Gunnarsson, Ciuchi, Sangiovanni, and Toschi}}]{Schafer:2013aa}
\bibinfo{author}{\bibfnamefont{T.}~\bibnamefont{Sch{\"a}fer}},
  \bibinfo{author}{\bibfnamefont{G.}~\bibnamefont{Rohringer}},
  \bibinfo{author}{\bibfnamefont{O.}~\bibnamefont{Gunnarsson}},
  \bibinfo{author}{\bibfnamefont{S.}~\bibnamefont{Ciuchi}},
  \bibinfo{author}{\bibfnamefont{G.}~\bibnamefont{Sangiovanni}},
  \bibnamefont{and} \bibinfo{author}{\bibfnamefont{A.}~\bibnamefont{Toschi}},
  \bibinfo{journal}{Physical Review Letters} \textbf{\bibinfo{volume}{110}},
  \bibinfo{pages}{246405} (\bibinfo{year}{2013}).

\bibitem[{\citenamefont{Chalupa et~al.}(2018)\citenamefont{Chalupa, Gunacker,
  Sch{\"a}fer, Held, and Toschi}}]{Chalupa:2018aa}
\bibinfo{author}{\bibfnamefont{P.}~\bibnamefont{Chalupa}},
  \bibinfo{author}{\bibfnamefont{P.}~\bibnamefont{Gunacker}},
  \bibinfo{author}{\bibfnamefont{T.}~\bibnamefont{Sch{\"a}fer}},
  \bibinfo{author}{\bibfnamefont{K.}~\bibnamefont{Held}}, \bibnamefont{and}
  \bibinfo{author}{\bibfnamefont{A.}~\bibnamefont{Toschi}},
  \bibinfo{journal}{Physical Review B} \textbf{\bibinfo{volume}{97}},
  \bibinfo{pages}{245136} (\bibinfo{year}{2018}).

\end{thebibliography}

\end{document}